\documentclass[prl,twocolumn,amsmath,amssymb,groupedaddress,showpacs]{revtex4-2}

\usepackage[T1]{fontenc} 
\usepackage{amsmath}
\usepackage{amssymb}
\usepackage{epsfig}

\usepackage{xcolor} 

\usepackage{ucs}
\usepackage{bbm}
\usepackage{bm}

\usepackage{orcidlink} 

\usepackage{hyperref}
\hypersetup{colorlinks=true, citecolor=blue, urlcolor=blue, linkcolor=blue}

\newcounter{myequation}
\makeatletter
\@addtoreset{equation}{myequation}
\makeatother
\newcounter{myfigure}
\makeatletter
\@addtoreset{figure}{myfigure}
\makeatother
\newcounter{mytable}
\makeatletter
\@addtoreset{table}{mytable}
\makeatother

\newcommand{\bra}[1]{\langle #1|}
\newcommand{\ket}[1]{|#1\rangle}
\newcommand{\ketbra}[1]{| #1\rangle \langle #1|}

\newcommand{\be}{\begin{equation}}
\newcommand{\ee}{\end{equation}}
\newcommand{\eea}{\end{eqnarray}}
\newcommand{\bea}{\begin{eqnarray}}

\newcommand{\va}[1]{\ensuremath{(\Delta#1)^2}}

\newcommand{\ex}[1]{\ensuremath{\langle{#1}\rangle}}
\newcommand{\exs}[1]{\ensuremath{\langle{#1}\rangle}}

\newcommand{\eins}{\mathbbm{1}}
\newcommand{\qed}{\ensuremath{\hfill \blacksquare}}

\newcommand{\kommentar}[1]{}
\newcommand{\trace}{{\rm Tr}}

\newcommand{\forget}[1]{}

\newcommand{\EQ}[1]{Eq.~\eqref{#1}}
\newcommand{\EQS}[1]{Eqs.~\eqref{#1}}
\newcommand{\EQL}[1]{Equation~\eqref{#1}}

\newcommand{\FIG}[1]{Fig.~\ref{#1}}
\newcommand{\REF}[1]{Ref.~\cite{#1}}
\newcommand{\REFS}[1]{Refs.~\cite{#1}}

\newcommand{\FQ}{\mathcal F_{Q}}

\newcommand{\CITESUPP}{\cite{Note2}}
\newcommand{\CITESUPPLABEL}{Note2}

\newcommand{\bipartite}{bipartite }

\begin{document}

\title{Activating hidden metrological usefulness}

\author{G\'eza T\'oth\,\orcidlink{0000-0002-9602-751X}}
\email{toth@alumni.nd.edu}
\homepage{http://www.gtoth.eu}
\affiliation{Department of Theoretical Physics, University of the Basque Country UPV/EHU, P.O. Box 644, E-48080 Bilbao, Spain}
\affiliation{Donostia International Physics Center (DIPC), P.O. Box 1072, E-20080 San Sebasti\'an, Spain}
\affiliation{IKERBASQUE, Basque Foundation for Science, E-48013 Bilbao, Spain}
\affiliation{Wigner Research Centre for Physics, Hungarian Academy of Sciences, P.O. Box 49, H-1525 Budapest, Hungary}
\author{Tam\'as V\'ertesi\,\orcidlink{0000-0003-4437-9414}}
\email{tvertesi@atomki.hu}
\affiliation{MTA Atomki Lend\"ulet Quantum Correlations Research Group, Institute for Nuclear Research, Hungarian Academy of Sciences, P.O. Box 51, H-4001 Debrecen, Hungary}
\author{Pawe{\l}  Horodecki}
\affiliation{International Centre for Theory of Quantum Technologies, University of Gda\'nsk, Wita Stwosza 63, 80-308 Gda\'nsk, Poland}
\affiliation{Faculty of Applied Physics and Mathematics, National Quantum Information Centre, Gda\'nsk University of Technology, Gabriela Narutowicza 11/12, 80-233 Gda\'nsk, Poland}
\author{Ryszard Horodecki}
\affiliation{International Centre for Theory of Quantum Technologies, University of Gda\'nsk, Wita Stwosza 63, 80-308 Gda\'nsk, Poland}
\affiliation{Institute of Theoretical Physics and Astrophysics, National Quantum Information Centre, Faculty of Mathematics, Physics and Informatics, University of Gda\'nsk, Wita Stwosza 57,80-308 Gda\'nsk, Poland}

\begin{abstract}
We consider \bipartite entangled states that cannot outperform separable states in any linear interferometer. Then, we show that these states can still be more useful metrologically than separable states  if several copies of the state are provided or an ancilla is added to the quantum system. We present a general method to find the local Hamiltonian for which a given quantum state performs the best compared to separable states. We obtain analytically the optimal Hamiltonian for some quantum states with a high symmetry. We show that all bipartite entangled pure states outperform separable states in metrology. Some potential applications of the results are also suggested.

\vspace{1em}
\noindent DOI: \href{https://doi.org/10.1103/PhysRevLett.125.020402}{10.1103/PhysRevLett.125.020402}
\end{abstract}

\date{\today}

\maketitle

Entanglement lies at the heart of quantum mechanics and plays an important role in quantum information theory \cite{Horodecki2009Quantum,*Guhne2009Entanglement,*Friis2019}. Recently, it has been realized that entanglement can be a useful resource in very general metrological tasks. By using entangled states it is possible to overcome the shot-noise limit, corresponding to classical interferometers, in the precision of parameter estimation \cite{Pezze2009Entanglement,Gessner2017Resolution,Hyllus2012Fisher,*Toth2012Multipartite,Lucke2011Twin,Krischek2011Useful,Strobel2014Fisher}. On the other hand, separable states, i.e., states without entanglement cannot overcome the classical limit. It has even been shown that quantum states with a very weak form of entanglement, called bound entanglement \cite{Horodecki1997Separability,Peres1996Separability,Horodecki1999Bound}, can also be metrologically useful in this sense \cite{Czekaj2015Quantum,Toth2018Quantum}. However, there are highly entangled states that are not useful for metrology \cite{Hyllus2010Not}.

\begin{figure}[b!]
\begin{minipage}[c]{3cm}
\includegraphics[scale=1.9]{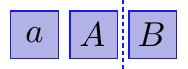}
\end{minipage}
\hskip1cm
\begin{minipage}[c]{3cm}
\includegraphics[scale=1.9]{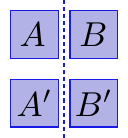}
\end{minipage}
\vskip0.2cm
\hskip0.5cm(a)\hskip3.2cm (b)
\caption{(a) An ancilla ("$a$") is added to bipartite state $\varrho_{AB}.$
(b) An additional copy or a different state is added to the state.
In both cases, a new bipartite state is obtained, where the two parties are separated by a dashed line.
} \label{fig:ancilla_twocopies}
\end{figure}

In what sense is metrological usefulness the property of the quantum state?  It is clear that, starting from many entangled quantum states that are not useful for metrology, with local operations and classical communication (LOCC) it is possible to distill singlets, which are metrologically useful. This finding is almost trivial, as metrological "uselessness" is not conserved by LOCC operations. On the other hand, in quantum metrology experiments  most LOCC operations are typically not possible. Here, we investigate how metrological usefulness can change in the two simplest cases very relevant in practice: We consider adding an ancilla to a single copy of the bipartite quantum state.  We also consider providing two copies of the state \footnote{In a general LOCC operation, large number of copies are used, and many rounds of classical communication take place. In our case {\it no classical communication} 
is needed, and in particular adding ancilla is a local operation (LO) which is an example of LOCC without classical communication (CC).}. These two scenarios follow the spirit in which the activation of bound entanglement and nonlocality has been studied \cite{Horodecki1999Bound,Nawarege2017Superadditivity,Activation2011Navascues,Palazuelos2012Superactivation} (see \FIG{fig:ancilla_twocopies}).

In this Letter, we show that some \bipartite entangled quantum states that are not useful in linear interferometers become useful in the cases mentioned above. These findings are quite surprising: including uncorrelated ancilla qubits can make a state metrologically useful.  To support our claims, we present a general method to find the {\it local} Hamiltonian for which a given \bipartite quantum state provides the largest gain compared to separable states. Note that this task is different, and in a sense more complex, than maximizing the quantum Fisher information. The reason is that by changing the Hamiltonian, the sensitivity achievable by separable states can also change.

{\it Quantum Fisher information.}---Before discussing our main results, we review some of the fundamental relations of quantum metrology. A basic metrological task in a {\it linear} interferometer is  estimating the small angle $\theta$ for a unitary dynamics $U_{\theta}=\exp(-i{\mathcal H}\theta),$ where the Hamiltonian is the sum of {\it local} terms. That is, all local terms act within the subsystem and there are no interactions between the subsystems.
In particular, for bipartite systems it is
\begin{equation}
{\mathcal H}={\mathcal H}_1+{\mathcal H}_2,\label{eq:bipartiteH}
\end{equation}
where ${\mathcal H}_n$ are single-subsystem operators. The precision is limited by the Cram\'er-Rao bound as \cite{Helstrom1976Quantum,*Holevo1982Probabilistic,*Braunstein1994Statistical,
*Petz2008Quantum,*Braunstein1996Generalized,Giovannetti2004Quantum-Enhanced,*Demkowicz-Dobrzanski2014Quantum,*Pezze2014Quantum,*Toth2014Quantum,Pezze2018Quantum,Paris2009QUANTUM}
\begin{equation} \label{eq:cramerrao} 
\va{\theta}\ge\frac 1 {m\FQ[\varrho,{\mathcal H}]},
\end{equation}
where $m$ is the number of indepedendent repetitions, and the quantum Fisher information, a central quantity in quantum metrology is defined by the formula \cite{Helstrom1976Quantum,*Holevo1982Probabilistic,*Braunstein1994Statistical,
*Petz2008Quantum,*Braunstein1996Generalized}
\begin{equation}
\label{eq:FQ}
\FQ[\varrho,{\mathcal H}]=2\sum_{k,l}\frac{(\lambda_{k}-\lambda_{l})^{2}}{\lambda_{k}+\lambda_{l}}\vert \langle k \vert {\mathcal H} \vert l \rangle \vert^{2}.
\end{equation}
Here, $\lambda_k$ and $\ket{k}$ are the eigenvalues and eigenvectors, respectively, of the density matrix $\varrho,$ which is used as a probe state for estimating $\theta.$

{\it Metrological usefulness of a quantum state.}---We call a quantum state metrologically useful, if it can outperform separable states in some metrological task, i.e., if
\begin{equation}
\FQ[\varrho,{\mathcal H}]>\max_{\varrho_{\rm sep}}\FQ[\varrho_{\rm sep},{\mathcal H}]=:\FQ^{({\rm sep})}({\mathcal H}).
\end{equation}
It is an intriguing task to find the operator ${\mathcal H},$ for which a given state performs the best compared to separable states. For that we define
the metrological gain compared to separable states by
\begin{equation}
g_{\mathcal H}(\varrho)=\FQ[\varrho,{\mathcal H}]/\FQ^{({\rm sep})}({\mathcal H}).\label{eq:FQ_over_FQsep}
\end{equation}
We are interested in the quantity
\begin{equation}
g(\varrho)=\max_{{\rm local } {\mathcal H}} g_{\mathcal H}(\varrho),\label{eq:glocalH}
\end{equation}
where a local Hamiltonian is just the sum of single system Hamiltonians as in \EQ{eq:bipartiteH}. The maximization task looks challenging since we have to maximize a fraction, where both the numerator and the denominator depend on the Hamiltonian.
(See the Supplemental Material for basic properties of the metrological gain \CITESUPP.)

{\it Maximally entangled state.}---As we have mentioned, it is a difficult task to obtain $g(\varrho)$ and the optimal local Hamiltonian for any $\varrho.$ 
As a first step, we consider the $d\times d$ maximally entangled state, which is defined as
\begin{equation}
\ket{\Psi^{({\rm me})}}=\frac 1 {\sqrt{d}} \sum_{k=1}^d\ket{k}\ket{k}.\label{eq:mestate}
\end{equation}
Due to the symmetry of the state, the optimal Hamiltonian can straightforwardly be obtained as
\begin{equation}
{\mathcal H}^{({\rm me})}=D\otimes\openone+\openone\otimes D,\label{eq:Hme}
\end{equation}
where the diagonal matrix $D$ is given as
\begin{equation}\label{eq:Ddef}
D={\rm diag}(+1,-1,+1,-1,...).
\end{equation}
 The details are given in the Supplemental Material \footnote{See Supplemental Material for additional results on metrology with isotropic states and Werner states, as well as for metrology with bipartite pure entangled states.  The Supplemental Material  includes \REF{Macieszczak2015Zeno, Apellaniz2015Detecting, Frowis2019Does,Toth2020InPreparation,Apellaniz2017Optimal,Sommers2004Statistical,Toth2006Two-setting,Popescu1992Generic,Oszmaniec2016Random,Greenberger1989Going,Krenn2016Automated,Uola2019Quantifying,Note4}.}. 
 \nocite{Macieszczak2015Zeno, Apellaniz2015Detecting,Frowis2019Does,Toth2020InPreparation,Apellaniz2017Optimal,Sommers2004Statistical,Toth2006Two-setting,Popescu1992Generic,Oszmaniec2016Random,Greenberger1989Going,Krenn2016Automated,Uola2019Quantifying,Note4}
\footnotetext[4]{T. Kraft, University of Siegen, Private communication (2019).}
For the $3\times 3$-case, 
we consider the noisy quantum state
\begin{equation}
\varrho^{(p)}_{AB}=(1-p)\ketbra{\Psi^{({\rm me})}}+p\openone/d^2, 
\end{equation}
which is useful if
\CITESUPP
\begin{equation}\label{eq:rphsi+}
p<\frac{25-\sqrt{177}}{32}\approx 0.3655.
\end{equation}
(See the Supplemental Material for the  definition of the related notion of robustness of metrological usefulness \CITESUPP.)

{\it Activation by an ancilla qubit.}---Now we consider the previous state, after a pure ancilla qubit is added
\begin{equation}
\varrho^{({\rm anc})}=\ketbra{0}_{a}\otimes\varrho_{AB}^{(p)}. \label{eq:anc}
\end{equation}
The setup is depicted in \FIG{fig:ancilla_twocopies}(a).
Then, with the operator
\begin{equation}\label{eq:XAAprime2}
\mathcal H^{({\rm anc})}=1.2 C_{aA}\otimes\openone_{B}+\openone_{aA}\otimes D_{B},
\end{equation}
where an operator acting on the ancilla and $A$ is
\begin{equation}
C_{aA}=\frac 9 {20} \left( 2\sigma_x+\sigma_z\right) _{a} \otimes\ketbra{0}_{a}
+  \openone_{a}\otimes(\ketbra{2}_{a}-\ketbra{1}_{a}),
\end{equation}
we have $g_{\mathcal H^{({\rm anc})}}(\varrho^{({\rm anc})})>1$ if $p<0.3752$ [c.f. \EQ{eq:rphsi+}]. Hence larger part of the noisy maximally entangled states are useful in the case with the ancilla.

{\it Activation by adding extra copies.}---We consider now two copies of the noisy $3\times3$ maximally entangled state
\begin{equation}
\varrho^{({\rm tc})}=\varrho^{(p)}_{AB}\otimes \varrho^{(p)}_{A'B'}.\label{eq:tc}
\end{equation}
The setup is shown in \FIG{fig:ancilla_twocopies}(b).
Then, with the two-copy operator
\begin{equation}\label{eq:XAAprime}
\mathcal H^{({\rm tc})}=D_{a}\otimes D_{A'} \otimes \openone_{B B'}+\openone_{A A'}\otimes D_{B}\otimes D_{B'},
\end{equation}
we have $g_{\mathcal H^{({\rm tc})}}(\varrho^{({\rm tc})})>1$ if $p<0.4164$  [c.f. \EQ{eq:rphsi+}]. Hence larger part of the noisy maximally entangled states are useful in the two-copy case, than with a single copy. So far we have studied the $3\times3$ case. For the $2\times2$-case, see the Supplemental Material \CITESUPP.

{\bf Observation 1.}---In summary, we have just shown that there are \bipartite states with the following properties. (i) They are  not more useful than separable states considering any local Hamiltonian. (ii) By adding an ancilla or two copies, they are more useful than separable states for some local Hamiltonian. For the case of an added ancilla, the new subsystems are now $aA$ and $B,$ and the Hamiltonian contains interactions between the ancilla $a$ and $A.$ In the two-copy case, the new subsystems are  $AA'$ and $BB',$ and the Hamiltonian contains interactions between $A$ and $A',$ and between $B$ and $B'.$ Note that in both cases, the extra interactions increase the metrological capabilities of separable states. Still, simple algebra shows  that in both cases the metrological gain can stay the same or can increase, but cannot decrease \CITESUPP.

So far, we exploited the symmetries of quantum states to obtain the Hamiltonian leading to the largest metrological gain. We now present a general method to compute $g(\varrho)$ numerically.

{\it Method for finding the optimal Hamiltonian.}---We need to maximize $\FQ[\varrho,{\mathcal H}]$ over ${\mathcal H}$ for a given  $\varrho$. However, since it is convex in $\mathcal H,$ maximizing it over $\mathcal H$ is a difficult task. Instead of the quantum Fisher information, let us consider the error propagation formula
\begin{equation}\label{eq:vartheta}
\va{\theta}_M= \frac{\va{M}}{\ex{i[M,{\mathcal H}]}^2},
\end{equation}
which provides a bound on the quantum Fisher information  \cite{Hotta2004Quantum,Escher2012Quantum,Frowis2015Tighter,Note2} 
\begin{equation}
\FQ[\varrho,{\mathcal H}]  \ge  1 / {\va{\theta}_M}. \label{eq:FQerrorprop}
\end{equation}
We will now minimize \EQ{eq:vartheta}.  

{\bf Observation 2}.---The error propagation formula given in \EQ{eq:vartheta} can be minimized over $\mathcal H$ for a given $M$ and $\varrho$ as follows.

{\it Proof.}
Simple algebra yields
\begin{equation}
\ex{i[M,{\mathcal H}]}={\rm Tr}(A_1 {\mathcal H}_1)+{\rm Tr}(A_2 {\mathcal H}_2),\label{eq:tracecomm}
\end{equation}
where $A_n={\rm Tr}_{\{1,2\}\backslash n}(i[\varrho,M] )$ are operators acting on a single subsytem.
Hence, we have to maximize \EQ{eq:tracecomm} over ${\mathcal H}_1$ and  ${\mathcal H}_2.$
We choose the constraints
\begin{equation}
\label{eq:cnHconst}
c_n \eins \pm \mathcal H_n \ge 0, 
\end{equation}
where $n=1,2$ and $c_n>0$ is some constant.
This way we make sure that
$
\sigma_{\min}(\mathcal H_n)\ge -c_n,
$
and
$
\sigma_{\max}(\mathcal H_n)\le+c_n,
$
for $n=1,2,$ where $\sigma_{\min}(X)$ and $\sigma_{\max}(X)$ denote the smallest and largest eigenvalues of $X.$ The optimal $\mathcal H_n$ is the one that maximizes ${\rm Tr}(A_n\mathcal H_n)$ under these constraints. It can straightforwardly be obtained as
\begin{equation}
\mathcal H_n^{({\rm opt})}=U_n \tilde D_n U_n^\dagger,\label{eq:Hopt}
\end{equation}
where the eigendecompisition of $A$ is given as $A_n=U_nD_nU_n^\dagger$ and $(\tilde D_n)_{k,k}=c_n{s}\bm{(}(D_n)_{k,k}\bm{)},$ where ${s}(x)=1$ if $x\ge 0,$ and $-1$ otherwise. Clearly, $\mathcal H_n^{({\rm opt})}$ has the same eigenvectors as $A_n$ and has only eigenvalues $+c_n$ and $-c_n.$ $\qed$

We already know how to optimize $\mathcal H$ for a given $M.$ However, how do we find the optimal $M?$ This can be done with the well-known formula for the symmetric logarithmic derivative~\cite{Paris2009QUANTUM}
\begin{equation}
\label{eq:SLD}
M_{\rm opt}=2i\sum_{k,l}\frac{\lambda_{k}-\lambda_{l}}{\lambda_{k}+\lambda_{l}} \vert k \rangle \langle l \vert \langle k \vert {\mathcal H} \vert l \rangle.
\end{equation}
For a given $\mathcal H,$ the error propagation formula given in \EQ{eq:vartheta} is minimized for $M=M_{\rm opt}$  \cite{Escher2012Quantum,\CITESUPPLABEL}.

{\it Iterative method.}---We can now construct the following procedure for minimizing \EQ{eq:vartheta}. First we choose a random $M.$ Then, repeat the following two steps.

(Step 1) Determine the optimal $\mathcal H$ for a given $M$ using Observation 2.

(Step 2) Determine the optimal $M$ for a given $\mathcal H$ using \EQ{eq:SLD}.

\noindent A see-saw procedure similar in spirit has been used to make the optimization of the metrological performance over density matrices in \REFS{Macieszczak2013Quantum,MacieszczakBayesian2014,Toth2018Quantum}.

After several iterations of the two steps above, we obtain the maximal quantum Fisher information over a certain set of Hamiltonians. Based on that, we can calculate the quantity
\begin{equation}\label{eq:gc1c2}
g_{c_1,c_2}(\varrho)=\max_{\mathcal H_1,\mathcal H_2} \frac{\FQ(\varrho,\mathcal H_1 \otimes \eins + \eins \otimes \mathcal H_2)}{\FQ^{({\rm sep})} (c_1,c_2)},
\end{equation}
where we assumed that $\mathcal H_n$ are constrained with \EQ{eq:cnHconst}.
The separable limit for Hamiltonians of the form \eqref{eq:bipartiteH} is \cite{Ciampini2016Quantum,Toth2018Quantum}
\begin{equation} \FQ^{({\rm sep})}(\mathcal H)=\sum_{n=1,2} [ \sigma_{\max}(\mathcal H_n)-\sigma_{\min}(\mathcal H_n) ]^2, \label{eq:seplim}\end{equation}
which leads to $
\FQ^{({\rm sep})} (c_1,c_2) = 4(c_1^2+c_2^2).$
Then,  the gain can be expressed as 
\begin{equation}
g(\varrho)=\max_{c_2}g_{c_1,c_2}(\varrho),
\end{equation}
where the optimization is only over $c_2$, and, without the loss of generality, we set $c_1=1.$ The optimal $c_2$ can be obtained from an analytical formula \CITESUPP. Hence we computed the maximum of the fraction, \eqref{eq:FQ_over_FQsep}, for local Hamiltonians.

We now stress the following. If we determine the optimal $\mathcal H$ for a given $M$ using Observation 2, the eigenvalues of the optimal $\mathcal H_n$ satisfying \EQ{eq:cnHconst} are $\pm c_n.$ Let us assume the contrary. Let us assume that for a state $\varrho$ and for given $c_1,c_2$ we know the optimal $\mathcal H_1$ and $\mathcal H_2,$ and $\mathcal H_n$  fulfill \EQ{eq:cnHconst}, but not all eigenvalues are $\pm c_n.$  We observe that $\ex{i[M,{\mathcal H}]}$ is a linear function of the eigenvalues of $\mathcal H_n,$ thus it takes its maximum at the eigenvalues corresponding to the boundary of the allowed region. Hence, we can always replace the eigenvalues of $\mathcal H_n$ by $\pm c_n$ such that $\ex{i[M,{\mathcal H}]}$ will not decrease, and $1/\va{\theta}_M$ will not decrease either. 

Using the numerical method above, we obtain a slightly larger value for the noise bounds of metrological usefulness for the state with an ancilla, \eqref{eq:anc}. $g(\varrho^{({\rm anc})})>1$ if $p<0.3941.$ The same is true for the case of the two copies of the noisy maximally entangled state, \eqref{eq:tc}. We obtain  $g(\varrho^{({\rm tc})})>1$ if $p<0.4170.$

For states with a high symmetry, such as isotropic states \cite{Horodecki1999Reduction,Horodecki1999General}, and Werner states \cite{Werner1989Quantum},
we obtained the optimal Hamiltonian analytically and determined the subset of these states that are metrologically useful \CITESUPP. We also used that to verify our numerical methods.

{\it Activation of a bound entangled state by a separable state.}---While bound entangled or non-distillable states \cite{Horodecki1997Separability,Peres1996Separability} are considered weakly entangled, they can share many properties with highly entangled states. For example, there are bound entangled states that can reach the Heisenberg scaling in metrological applications \cite{Czekaj2015Quantum}.
It has also been shown that bipartite bound entangled states, which have a positive semidefinite partial transposition (PPT), can be useful for metrology  \cite{Toth2018Quantum}.
Moreover, bipartite PPT entangled states can even have a high Schmidt-rank \cite{Huber2018High}.

Let us now consider a PPT entangled state $\varrho^{({\rm PPT)}}_{AB}$ that is not useful for quantum metrology.
Then, we look for a separable state  $\varrho^{({\rm sep})}$  such that $\varrho^{({\rm PPT)}}_{AB}\otimes\varrho^{({\rm sep)}}_{A'B'}$ becomes useful.
Hence, in this case we have to optimize not only over $\mathcal H,$ $M,$ but also over the  separable state. Simple convexity arguments show that the maxiumum is taken when we have a pure product state, $\varrho^{({\rm sep})}_{A'B'}=\varrho_{A'}^{({\rm anc})}\otimes\varrho_{B'}^{({\rm anc})},$ which corresponds to two ancillas at the two parties. In fact, even a single ancilla qubit is sufficient for activation.

{\it Activation of a PPT entangled state by an ancilla qubit.}---We now consider a PPT entangled state, that is not useful metrologically, and $g(\varrho_{AB})=1$. However, with an ancilla it becomes useful, $g(\varrho_{(aA)(B)})>1$. We show here examples for $d\times d$ dimensional PPT states found in Ref.~\cite{Toth2018Quantum} for odd dimensions $d$ up to $d\le 11$. See Table~\ref{tab:FQactivate_ppt} for the numerical results.

Note that here we fixed $c_i=1$ for the coefficients of the local Hamiltonians $\mathcal{H}_i$, $i=1,2$. However, numerics suggests that optimization over $c_i$ does not help to increase $g$ in the case of two ancillas (last column), due to the permutational symmetry of the states. Optimization over $c_i$ helps only marginally in the case of one ancilla (third column). For instance, in the case of $d=7$, the $g$ value raises from $1.0096$ (corresponding to $c_2=1$) to $1.0098$ (corresponding to $c_2\simeq 1.034$) if we optimize over $c_2$.

\begin{table}[t!]
\begin{center}
\begin{tabular}{|l|c|c|c|}
\hline
$d$ & $p^*$ & $\begin{array}{c} \text{Gain with} \\ \text{one ancilla} \end{array}$& $\begin{array}{c} \text{Gain with} \\ \text{two ancillas} \end{array}$
\\
\hline
\hline
$3$ & $0.0006$ & $1.0007$& $1.0011$\\
\hline
$5$ & $0.0960$ & $1.0094$& $1.0190$\\
\hline
$7$ & $0.1377$ & $1.0096$& $1.0195$\\
\hline
$9$ & $0.1631$ & $1.0090$& {1.0181}\\
\hline
$11$ & $0.1807$ & $1.0081$& {1.0165}\\
\hline
\end{tabular}
\end{center}
\caption{
Activation of the metrological usefulness found numerically in two-qudit systems.
(First column) Local dimension $d$, where $d$ is odd. For even $d$ up to $d\le 11$, we did not find activation in the examples of PPT two-qudit states considered.
(Second column) White noise fractions of $p^*$  added to the PPT states given by Ref.~\cite{Toth2018Quantum} such that $g_{1,1}(\varrho_{AB})=1.0000$, that is, they are not useful metrologically.
(Third column) Metrological gain after an ancilla is added to Alice's system, $g_{1,1}(\varrho_{(aA)(B)}).$ The states become useful as demonstrated by $g_{1,1}(\varrho_{(aA)(B)})>1.$
(Fourth column) Metrological gain after a further ancilla is added to Bob's system, $g_{1,1}(\varrho_{(aA)(Bb)}).$ The state becomes even more useful metrologically.
}
\label{tab:FQactivate_ppt}
\end{table}

{\it Entanglement detection.}---Our method can be used for entanglement detection. It identifies the Hamiltonians with which a given quantum state performs better than separable states and hence it is detected as entangled. If we add ancillas or extra copies of the quantum state, the criterion can be even more powerful.

{\it Random states.}---We can use our method to determine the distribution of metrological usefulness of random pure or mixed states of a given size.  For instance, for $3\times3$ systems, pure states typically are close to be maximally useful, while this is not the case if we look for the usefulness with respect to a given Hamiltonian. For the numerical result, please see the Supplemental Material \CITESUPP.

{\it Usefulness of entangled bipartite pure states.}---Next we will consider the usefulness of bipartite pure states.

{\bf Observation 3.}---All entangled bipartite pure states are metrologically useful. (For the two-qubit case, see \REF{Hyllus2010Not}.) 

{\it Proof.}---Let us consider a pure state with a Schmidt decomposition
\begin{equation}
\ket{\Psi}=\sum_{k=1}^{s} \sigma_k \ket{k}_{a}\ket{k}_{B},
\label{schmidt}
\end{equation}
where $s$ is the Schmidt number, and the real
positive $\sigma_k$ Schmidt coefficients are in a descending
order.
We define
 \begin{equation}
\mathcal H_{a}=\sum_{n=1,3,5,...,\tilde{s}-1}\ketbra{+}_{{\rm A},n,n+1}-\ketbra{-}_{{\rm A},n,n+1},
\label{eq:HA}
\end{equation}
where $\tilde{s}$ is the largest even number for which $\tilde{s}\le s,$ and
\begin{equation}
\ket{\pm}_{{\rm A},n,n+1}=(\ket{n}_{a}\pm\ket{n+1}_{a})/\sqrt{2}.
\end{equation}
We define $\mathcal H_{B}$ in a similar manner.
We also
define the collective Hamiltonian
\begin{equation}\label{eq:HAB}
\mathcal H_{AB}=\mathcal H_{a}\otimes\openone+\openone \otimes \mathcal
H_{B}.
\end{equation} Then, we have $\ex{\mathcal H_{AB}}_{\Psi}=0.$
Direct calculation yields
 \begin{equation}
 \FQ[\ket{\Psi},\mathcal H_{AB}]=4\va{\mathcal H_{AB}}_\Psi=8\sum_{n=1,3,5,...,\tilde s-1}(\sigma_n+\sigma_{n+1})^2,
\label{evenS}
 \end{equation}
which is larger than the separable bound, $\FQ^{({\rm sep})}=8,$ whenever the Schmidt rank is larger than $1.$  For even $s,$ this can be seen noting that
 \bea
 \FQ[\ket{\Psi},\mathcal H_{AB}]
&>& 8 \sum_{n=1}^s \sigma_n^2 
\label{eq:FQbound}
 \end{eqnarray}
 holds, where we used \EQ{evenS} to evaluate the left-hand side of \EQ{eq:FQbound}, and we also took into account that $\sigma_n> 0$ for $n=1,2,3,...,$ and $\sum_{n=1}^s \sigma_n^2=1.$ For odd $s,$ we need that
 \begin{equation}
\FQ[\ket{\Psi},\mathcal H_{AB}] \ge 8 \left( \sum_{n=1}^{s-1} \sigma_n^2 + 2\sigma_1\sigma_2 \right)> 8 \sum_{n=1}^s \sigma_n^2  
 \end{equation} holds, where we used that $\sigma_1\sigma_2>\sigma_s^2.$ $\qed$
 
 We can even consider several copies of a quantum state. In the Supplemental Material, we prove that for infinite number of copies of entangled pure quantum states the metrological gain is maximal \CITESUPP.

{\it Conclusions.---}We showed that  entangled quantum states that cannot outperform separable states in any linear interferometer  can still be more useful than separable states, if several copies of them are considered or an ancilla is added to the system.  This is a surprising result which shows that the relationship between quantum metrology and the structure of quantum states requires further study. We presented a method to find the Hamiltonian for carrying out metrology in a linear interferometer with a given quantum state that provides the largest gain compared to the precision achievable by separable states. In the Letter we considered  bipartite problems, thus it would be important to extend this approach to multipartite systems and examine the scaling of the metrological gain for noisy quantum states. It would be also interesting to look for application in entanglement detection \cite{Horodecki2009Quantum,*Guhne2009Entanglement,*Friis2019}, and witnessing the dimension of quantum systems \cite{Bowles2014Certifying,Brunner2008Testing,Gallego2010Device,Navascues2015Bounding}, where the results of the preliminary  analysis  seem to be promising. (See the Supplemental Material \CITESUPP.)

We thank I.~Apellaniz, D.~Gross, O.~G\"uhne, S.~Imai, M.~Kleinmann, J. Ko\l ody\'nski, T.~Kraft, J.~Siewert, and G.~Vitagliano for discussions. We acknowledge the support of the  EU (ERC Starting Grant 258647/GEDENTQOPT, COST Action CA15220, QuantERA CEBBEC, QuantERA eDICT), the Spanish MCIU (Grant No. PCI2018-092896), the Spanish Ministry of Science, Innovation and Universities and the European Regional Development Fund FEDER through Grant No. PGC2018-101355-B-I00 (MCIU/AEI/FEDER, EU), the Basque Government (Grant No. IT986-16), and the National Research, Development and Innovation Office NKFIH (Grants No.  K124351, No. KH129601, No. KH125096 and No. 2019-2.1.7-ERA-NET-2020-00003).  We also acknowledge support by the Foundation for Polish Science through IRAP project co-financed by the EU within the Smart Growth Operational Programme (Contract No. 2018/MAB/5). G.T. thanks a  Bessel Research Award of the Humboldt Foundation.

\bibliography{Bibliography2}

\clearpage

\renewcommand{\thefigure}{S\arabic{figure}}
\renewcommand{\thetable}{S\arabic{table}}
\renewcommand{\theequation}{S\arabic{equation}}
\stepcounter{myfigure}

\stepcounter{mytable}

\stepcounter{myequation}
\setcounter{page}{1}
\thispagestyle{empty}

\onecolumngrid


\begin{center}
{\large \bf Supplemental Material for \\``Activating hidden metrological usefulness''}

\bigskip
G\'eza T\'oth

{\it  \small Department of Theoretical Physics,
University of the Basque Country
UPV/EHU, P.O. Box 644, E-48080 Bilbao, Spain}

{\it  \small Donostia International Physics Center (DIPC), 
P.O. Box 1072,
E-20080 San Sebasti\'an, Spain}

{\it  \small IKERBASQUE, Basque Foundation for Science,
E-48013 Bilbao, Spain}

{\it  \small Wigner Research Centre for Physics, Hungarian Academy of Sciences, P.O. Box 49, H-1525 Budapest, Hungary}

\bigskip
Tam\'as V\'ertesi

{\it  \small MTA Atomki Lend\"ulet Quantum Correlations Research Group, Institute for Nuclear Research, \\ Hungarian Academy of Sciences, P.O. Box 51, H-4001 Debrecen, Hungary}

\bigskip
Pawe{\l}  Horodecki

{\it  \small International Centre for Theory of Quantum Technologies,
University of Gda\'nsk, Wita Stwosza 63, 80-308 Gda\'nsk, Poland}

{\it  \small Faculty of Applied Physics and Mathematics, National Quantum Information Centre, Gda\'nsk University of Technology, Gabriela Narutowicza 11/12, 80-233 Gda\'nsk, Poland}

\bigskip
Ryszard Horodecki

{\it \small International Centre for Theory of Quantum Technologies,
University of Gda\'nsk, Wita Stwosza 63, 80-308 Gda\'nsk, Poland}

{\it \small Institute of Theoretical Physics and Astrophysics, National Quantum Information Centre, Faculty of Mathematics, Physics and Informatics, University of Gda\'nsk, Wita Stwosza 57,80-308 Gda\'nsk, Poland}

(Dated: \today)

\medskip
\medskip

\parbox[b][1cm][t]{0.85\textwidth}{\quad
The Supplemental Material contains some additional results. We present some properties of the metrological gain. We discuss the relation between the error propagation formula and the quantum Fisher information. We present some details of the optimization over the $c_2$ parameter of the Hamiltonian. We calculate the optimal Hamiltonian analytically for isotropic states and Werner states. We present concrete calculations for metrology with two-qubit singlets and ancillas. We show how to use our formulas to bound the metrological usefulness by a single operator expectation value. We consider metrology with multi-particle states, if some particles are united into a single party. We consider metrology with an infinite number of copies of arbitrary entangled pure states. We present an alternative optimization over local Hamiltonians. We present numerical results concerning metrology with random pure and mixed states. We determine the maximum achievable precision in a multiparticle system.  We define the robustness of metrological usefulness.  We show how to witness the dimension of a quantum state based on quantum metrology.
}
\bigskip
\bigskip
\bigskip
\bigskip
\bigskip
\bigskip
\bigskip
\bigskip
\bigskip
\end{center}

\bigskip

\twocolumngrid

\section {Properties of the metrological gain in multipartite systems}

We consider the question, how the metrological gain defined in \EQ{eq:glocalH} behaves if we 
add an ancilla to the subsystem or provide an additional state, as depicted by \FIG{fig:ancilla_twocopies}. We will now show that it cannot decrease in neither of these cases.
We will also show that the metrological gain is convex.

(i) Let us see first adding an ancilla "$a$" to the system $AB.$ For the gain, we have
\begin{eqnarray}
&&g(\varrho_{AB})=g_{\mathcal H_{\rm opt}}(\varrho_{AB})\nonumber\\
&&\quad=g_{\mathcal H_{\rm opt}' }(\ketbra{0}_a\otimes \varrho_{AB})
\le g(\ketbra{0}_a\otimes \varrho_{AB}),\label{eq:metgain}
\end{eqnarray}
where a Hamiltonian for the $aAB$ system is given as
\be
\mathcal H_{\rm opt}' = \openone_a\otimes (\mathcal  H_{\rm opt})_{AB}.
\ee
Here, $\mathcal H_{\rm opt}$ is the local Hamiltonian acting on $AB$ for which the gain is the largest. 
The second equality in \EQ{eq:metgain} holds, since the quantum Fisher information has the property
\be
\FQ[\varrho_1\otimes \varrho_2,\mathcal H_1 \otimes \openone + \openone \otimes \mathcal H_2]=
\FQ[\varrho_1,\mathcal H_1]+\FQ[\varrho_2,\mathcal H_2].
\ee
For $\mathcal H_1=0,$ we have the special case 
\be
\FQ[\varrho_1\otimes \varrho_2,\openone \otimes \mathcal H_2]=\FQ[\varrho_2,\mathcal H_2].
\ee
The inequality in \EQ{eq:metgain} holds, since in the extended system there might be a Hamiltonian with a gain larger than that of 
$\mathcal H_{\rm opt}'.$ In other words, for any $\mathcal H$ and any  $\varrho$, $g_{\mathcal H}(\varrho) \le g (\varrho)$ holds. 

(ii) For an additional copy of a state, analogously, we have 
\begin{eqnarray}
&&g(\varrho_{AB})=g_{\mathcal H_{\rm opt}}(\varrho_{AB})\nonumber\\
&&\quad=g_{\mathcal H_{\rm opt}'' }(\varrho_{AB}\otimes \sigma_{A'B'})
\le g(\varrho_{AB}\otimes \sigma_{A'B'}),\label{eq:metgain2}
\end{eqnarray}
where a Hamiltonian for the $ABA'B'$ system is given as
\be
\mathcal H_{\rm opt}'' = (\mathcal H_{\rm opt})_{AB} \otimes \openone_{A'B'}.
\ee
Here, $\sigma_{A'B'}$ is  the additional state provided.
In the special case of two copies we have $\sigma=\varrho.$
If we replace the role of $\varrho_{AB}$ and $\sigma_{A'B'}$ in \EQ{eq:metgain2},
we arrive at
\be
g(\sigma_{AB}) \le g( \sigma_{AB} \otimes \varrho_{A'B'}).\label{eq:metgain3}
\ee
From \EQS{eq:metgain2} and \eqref{eq:metgain3}, after trivial relabelling of the parties follows
\be
g(\varrho_{AB}\otimes \sigma_{A'B'})\ge \max[g(\varrho_{AB}),g(\sigma_{A'B'})],
\ee
where $\max(a,b)$ denotes the maximum of $a$ and $b.$

(iii) The metrological gain is convex under mixing as can be seen from the series of inequalities
\begin{eqnarray}
g(p\varrho+(1-p)\sigma)&=&g_{\mathcal H_{\rm opt}}(p\varrho+(1-p)\sigma)\nonumber\\
&\le& p g_{\mathcal H_{\rm opt}}(\varrho)+(1-p) g_{\mathcal H_{\rm opt}}(\sigma)\nonumber\\
&\le& p g(\varrho)+(1-p) g(\sigma),
\end{eqnarray}
where $0\le p\le 1.$
Here, $\mathcal H_{\rm opt}$ is the Hamiltonian acting on $p\varrho+(1-p)\sigma$ for which the gain is the largest. The first inequality is due to the convexity of the quantum Fisher information. The second inequality is due to the fact, that in general for any $\mathcal H$ and any $\varrho$, $g_{\mathcal H}(\varrho) \le g (\varrho)$ holds.

\section {Relation between the error-propagation formula and the Quantum Fisher information}

\EQL{eq:FQerrorprop} has been described from various point of views in \REFS{Hotta2004Quantum,Escher2012Quantum,Frowis2015Tighter}.
These ideas have been used in \REFS{Pezze2009Entanglement,Macieszczak2015Zeno, Apellaniz2015Detecting,Frowis2019Does}.
Related ideas have also been used in \REFS{Macieszczak2013Quantum,MacieszczakBayesian2014} for the optimization of the quantum Fisher information.

For completeness, now we prove \EQ{eq:FQerrorprop} very briefly.
Let us consider the uncertainty relation \cite{Frowis2015Tighter,Toth2020InPreparation}
\be
\va{A}_\varrho \FQ[\varrho,B]\ge \ex{i[A,B]}_\varrho^2,\label{eq:varFQ}
\ee
where $\varrho$ is a quantum state, and $A$ and $B$ are observables. 
\REF{Frowis2015Tighter} stresses the fact that \EQ{eq:varFQ} is just a strengthening of the Heisenberg uncertainty relation.
Then, making the substitutions in \EQ{eq:varFQ}  that $B=\mathcal H$, $A= M$, we find that 
\be
\va{\theta}_M\equiv\frac{\va{M}}{\ex{i[M,{\mathcal H}]}^2} \ge 1/\FQ[\varrho,\mathcal H]\label{eq:varFQ2}
\ee
holds, where the left hand-side is just the error propagation formula. We now show that setting $M$ to the symmetric logarithmic derivative $M_{\rm opt}$ given in \EQ{eq:SLD} the inequality in 
\EQ{eq:varFQ2} is saturated. This can be proved using the identities $\trace(M_{\rm opt}^2\varrho)=\FQ[\varrho,\mathcal H],$ $\trace(M_{\rm opt}\varrho)=0,$
$\ex{i[M_{\rm opt},{\mathcal H}]}=\trace(M_{\rm opt}^2\varrho).$

Note that \EQ{eq:FQerrorprop} is different from the Cram\'er-Rao bound, \eqref{eq:cramerrao}.
The relation between the precision $\va{\theta}$ for some estimator and the error propagation formula $\va{\theta}_M$ is not trivial.
For any estimator 
\be
\va{\theta} \ge  \frac{1}{m} \va{\theta}_{M=M_{\rm opt}}\label{eq:varFQ2b}
\ee
holds. In the limit of large number of repetitions $m,$ and if certain further conditions are fulfilled,  \EQ{eq:varFQ2b} can be saturated by the best estimator.
Then, such a $\va{\theta}$ would also saturate the Cram\'er-Rao bound, \eqref{eq:cramerrao} \cite{Pezze2018Quantum}.

\section {Analysis of the optimization method}

The maximization of the error propagation formula can be expressed using a variational formulation as \cite{Macieszczak2013Quantum}
\begin{eqnarray}
&&\max_{{\mathcal H}}\max_M 1/{\va{\theta}_M}\nonumber\\
&&\quad\quad=\max_{\mathcal H} \max_{M} {\ex{i[M,{\mathcal H}]}^2}/{\va{M} }\nonumber\\
&&\quad\quad=\max_{\mathcal H} \max_{M} {\ex{i[M,{\mathcal H}]}^2}/{\ex{M^2} }\nonumber\\
&&\quad\quad=\max_{\mathcal H} \max_M \max_\alpha \{ -\alpha^2 \ex{M^2} + 2 \alpha \ex{i[M,{\mathcal H}]}  \}\nonumber\\
&&\quad\quad=\max_{\mathcal H} \max_{M'} \{ -\ex{(M')^2} + 2 \ex{i[M',{\mathcal H}]} \},\label{eq:fmin}
\end{eqnarray}
where $M'$ takes the role of $\alpha M.$  Then, the function is concave in $M'$ and linear in $\mathcal H,$ and the two-step see-saw algorithm we have described will find better and better Hamiltonians. However, the function in \EQ{eq:fmin} is not strictly concave in $(\mathcal H,M').$ Hence, our iterative numerical procedure will always lead to Hamiltonians with an increasing quantum Fisher information, however, it is not guaranteed to find a global optimum. Based on extensive numerical experience, for a mixed state in bipartite systems of dimension $3\times3$ the algorithm converges very fast, and from 10 trials at least 2-3, typically more will lead to the global optimum. The 10 trials of 100 steps can take 5 seconds on a state of the art laptop computer. For larger systems, it is worth to make many trials for few steps, and continue the best one for many steps. 

We can understand the expression better as follows.
If we subtract a term $4\ex{\mathcal H^2}$ from the expression appearing on the right-hand side of \EQ{eq:fmin}, then we will arrive at
\be
-\ex{ZZ^\dagger },\label{eq:Z}
\ee
where the non-Hermitian matrix is defined as
\be
Z=M'+i2\mathcal H.
\ee \EQL{eq:Z} is  clearly concave in $(\mathcal H,M')$ but a maximization will converge to $(\mathcal H,M')=0.$ The maximization in \EQ{eq:fmin} is equivalent to  maximizing \EQ{eq:Z} with a quadratic equality constraint $\ex{\mathcal H^2}=c,$ where $c$ is some constant.
We can maximize \EQ{eq:Z} for a range of $c$ values, and the largest of these maxima will be the global maximum.

\section{Efficient optimization over $c_2.$}

Let us define  $\tilde{ {\mathcal H}}_k=  {\mathcal H}_k/c_k.$
Based on \EQ{eq:cnHconst},
\be
-\openone \le \tilde{ {\mathcal H}}_k \le \openone
\ee
hold. Then, the Hamiltonian, \eqref{eq:bipartiteH}, becomes 
\be
\mathcal H=c_1\tilde{ {\mathcal H}}_1+c_2\tilde{ {\mathcal H}}_2.\label{eq:Hc1c2}
\ee
In this section, we show how to optimize the metrological performance for Hamiltonians of the form \eqref{eq:Hc1c2}. This will mean an optimization over $c_2,$ while $c_1$ can be taken to be $1.$ 

For a such $\tilde{ {\mathcal H}}_k$ Hamiltonians, the expression in \EQ{eq:tracecomm} can be written as
\begin{eqnarray}
\ex{i[M,{\mathcal H}]}
=c_1{\rm Tr}(A_1 \tilde{ {\mathcal H}}_1)+c_2{\rm Tr}(A_2 \tilde{ {\mathcal H}}_2),
\end{eqnarray}
where $A_n={\rm Tr}_{\{1,2\}\backslash n}(i[\varrho,M] ).$
Then, in order to maximize $\sqrt{\va{\theta}_M/\FQ^{({\rm sep)}}}$, we need to calculate
\be
\max_{c_1,c_2} \frac{c_1{\rm Tr}(A_1 \tilde{ {\mathcal H}}_1)+c_2{\rm Tr}(A_2 \tilde{ {\mathcal H}}_2)}{4\sqrt{c_1^2+c_2^2}}.
\ee
The optimal value is at
\begin{eqnarray}
\frac{c_2}{c_1}=\frac{{\rm Tr}(A_2 \tilde{ {\mathcal H}}_2)}{{\rm Tr}(A_1 \tilde{ {\mathcal H}}_1)}.\label{eq:c1c2}
\end{eqnarray}
Without the loss of generality, we set $c_1=1,$ then $c_2$ can be obtained from \EQ{eq:c1c2}.

One can add a third step to the two-step procedure of the paper, in which $c_2$ is updated according to the formula \EQ{eq:c1c2}. For a smoother convergence, one can change $c_2$ not abruptly, but only by a small value changing it in the direction of the value suggested by  \EQ{eq:c1c2}.

\section{Metrology with isotropic states}
\label{app:met_iso}

We will now consider quantum metrology with isotropic states, which are defined as \cite{Horodecki1999Reduction}
\be\label{eq:rhop}
\varrho_{p}=pP_d^{(+)}+(1-p)\frac{\openone}{d^2},
\ee
where $P_d^{(+)}$ is a projector to the maximally entangled state $\ket{\Psi^{({\rm me})}}$ defined in \EQ{eq:mestate}. 

We consider a Hamiltonian of the form
\be
\mathcal{H}_{\rm coll}=\mathcal{H}_1 \otimes \mathbbm{1}+\mathbbm{1} \otimes \mathcal{H}_2.\label{eq:Ageneral}
\ee
The subscript "coll" indicates that the Hamiltonian acts on both subsystems, in contrast to  $\mathcal{H}_1$ and $\mathcal{H}_2$ that act only on one of the subsystems.
The Hamiltonian is local, since it does not contain interactions terms.

Isotropic states are invariant under transformations of the type
\be
U\otimes U^*,\label{eq:UUstar}
\ee
where $U$ is a single-qudit unitary and "$^*$" denotes element-wise conjugation. Hence, isotropic states are invariant under the Hamiltonian
\be
\mathcal H_{\rm inv}^{({\rm iso})}(\mathcal{H})=\mathcal K  \otimes \mathbbm{1}-\mathbbm{1} \otimes \mathcal K^*,\label{eq:Aantisym}
\ee
where $\mathcal K$ is a Hermitian operator.

{\bf Observation S1.}---For short times, the action of the Hamiltonian $\mathcal{H}_{\rm coll}$ given in \EQ{eq:Ageneral} is the same as the action of
\be
\mathcal H_{\rm coll}^{({\rm iso})}(\mathcal{H}^{({\rm iso})})=\mathcal{H}^{({\rm iso})} \otimes \mathbbm{1}+\mathbbm{1} \otimes (\mathcal{H}^{({\rm iso})})^*,\label{eq:Asym}
\ee
where the single party Hamiltonian is defined as
\be
\mathcal{H}^{({\rm iso})}=(\mathcal{H}_1+\mathcal{H}_2^*)/2. \label{eq:Hav}
\ee

{\it Proof.} Let us define
\be
\Delta^{({\rm iso})} =(\mathcal H_2^*-\mathcal H_1)/{2}.\label{eq:Delta}
\ee
In the rest of the section, we omit the superscript "iso" in $\mathcal H_{\rm inv}^{({\rm iso})}, \mathcal{H}^{({\rm iso})}, \Delta^{({\rm iso})}.$

Then, simple algebra shows that
\be
\mathcal{H}_{\rm coll}+\mathcal H_{\rm inv}\left(\Delta\right)=\mathcal H_{\rm coll}^{({\rm iso})}.
\ee
Hence, for small $t$
\be
e^{-i \mathcal{H}_{\rm coll}t}e^{-i \mathcal H_{\rm inv}(\Delta) t}\approx e^{-i \mathcal H_{\rm coll}^{({\rm iso})}(\mathcal{H})t}\label{eq:approx}
\ee
holds. The isotropic state is invariant under the action of $\mathcal H_{\rm inv}(\Delta),$
since the corresponding unitary is of the form given in \EQ{eq:UUstar}.
Hence, the action of $\mathcal{H}_{\rm coll}$ is the same as the action of $H_{\rm coll}^{({\rm iso})}(\mathcal{H})$ for small $t$.  $\qed.$

Note that in the quantum metrology problems we consider we always estimate the parameter $t$ around $t=0$ assuming that it is small.
Hence, the approximate equality in \EQ{eq:approx} is sufficient.

{\bf Observation S2.}---Replacing the evolution by $\mathcal{H}_{\rm coll}$ given in \EQ{eq:Ageneral} by the evolution by $\mathcal{H}_{\rm coll}^{({\rm iso})}$ given in \EQ{eq:Asym}
does not decrease the metrological gain. Hence, when looking for the Hamiltonian with the largest metrological gain, it is sufficient to look for Hamiltonians of the form \eqref{eq:Asym}.

{\it Proof.}  When the evolution by $\mathcal{H}_{\rm coll}$ given in \EQ{eq:Ageneral} is replaced by the evolution by $\mathcal{H}_{\rm coll}^{({\rm iso})}$  then the quantum Fisher information does not change, while $\FQ^{({\rm sep})}$ does not increase. The latter can be seen as follows. Let us define
\be
f(X)= [ \sigma_{\max}(X)-\sigma_{\min}(X) ]^2,
\ee
where $X$ is some matrix. Then, based on \EQ{eq:seplim}, $\FQ^{({\rm sep})}(\mathcal H_{\rm coll})=f(\mathcal H_1)+f(\mathcal H_2)$ holds. On the other hand, we have
$\FQ^{({\rm sep})}(\mathcal H_{\rm coll}^{({\rm iso})})=2f(\mathcal H).$
Knowing that $f$ is matrix convex, we obtain that
\be
\FQ^{({\rm sep})}(\mathcal H_{\rm coll}^{({\rm iso})})\le \FQ^{({\rm sep})}(\mathcal H_{\rm coll}).
\ee$\qed$

We will now use that for a pure state mixed with white noise it is possible to obtain a closed formula for the quantum Fisher information for any operator $A$ as a function of $p$ as \cite{Toth2012Multipartite,Hyllus2012Fisher}
\be
\FQ[\varrho_{p},A]=\frac{p^2}{p+2(1-p)d^{-2}}4\va{A}_{\Psi^{({\rm me})}},\label{eq:FQ_iso1}
\ee
where $\varrho_{p}$ given in \EQ{eq:rhop}.
Let us simplify \EQ{eq:FQ_iso1}. For the case of $A=\mathcal H^{({\rm iso})}_{\rm coll},$ we can rewrite the variance as
\begin{eqnarray}
\va{\mathcal H^{({\rm iso})}_{\rm coll}}_{\Psi^{({\rm me})}}&=&
2\frac{\trace(\mathcal H^2)}{d}+2\ex{\mathcal H\otimes \mathcal H^*}_{\Psi^{({\rm me})}}-4\frac{\trace(\mathcal H)^2}{d^2},\nonumber\\
\end{eqnarray}
where we used that for the reduced state of $\ket{\Psi^{({\rm me})}}$ we have $\rho_{\rm red1}=\rho_{\rm red2}=\openone/d.$
Next, we use the fact that
\be
\ex{\mathcal H\otimes \mathcal H^*}_{\Psi^{({\rm me})}}=\frac{1}{d}\trace(\mathcal H^2)
\ee
holds. Hence, for the quantum Fisher information we obtain
\be
\FQ[\varrho_{p},\mathcal H^{({\rm iso})}_{\rm coll}]=\frac{16p^2}{pd^2+2(1-p)}\left[d\trace(\mathcal H^2)-\trace(\mathcal H)^2\right].\label{eq:FQ_iso2}
\ee
Based on \EQ{eq:FQ_iso2} and on \EQ{eq:seplim}, the metrological gain for a given Hamiltonian $\mathcal H^{({\rm iso})}_{\rm coll}$ is obtained as
\be \label{eq:performance_iso}
g(\varrho_p,\mathcal H^{({\rm iso})}_{\rm coll})=\frac{16p^2}{pd^2+2(1-p)}r(\mathcal H),
\ee
where $r(\mathcal H)$ is defined as
\begin{eqnarray}
r(\mathcal H)&=&\frac{[d \sum_k h_k^2-(\sum_k h_k)^2]}{2(h_{\max}-h_{\min})^2},\label{eq:RH2}
\end{eqnarray}
and $h_k$ denote the eigenvalues of $\mathcal H.$

Let us now consider the metrological gain for the isotropic state for various Hamiltonians.

{\bf Observation S3.}---Isotropic states have the best metrological performance with respect to separable states with the Hamiltonian given by
\be
\mathcal H_{\rm best}={\rm diag}(+1,-1,+1,-1,...).
\label{eq:Hbest_anyd}
\ee
Based on \EQ{eq:FQ}, the corresponding metrological performance is described by
\be
g(\varrho_p,\mathcal H^{({\rm iso})}_{\rm coll}({\mathcal H}_{\rm best}))=\frac{2p^2[d^2-\alpha]}{pd^2+2(1-p)},
\label{eq:iso_ratioFQ_FQsep_best}
\ee
where $\alpha$ is defined as
\be\label{eq:alpha_evenodd}
\alpha=\bigg\{
\begin{array}{cc}
0& \text{ for even } d, \\
1& \text{ for odd } d.
\end{array}
\ee
No other Hamiltonian $\mathcal H$ corresponds to a better performance. 

\EQL{eq:iso_ratioFQ_FQsep_best} is maximal for $p=1$ and has the value
\be
g(\varrho_p,\mathcal H^{({\rm iso})}_{\rm coll}({\mathcal H}_{\rm best}))=2\frac{d^2-\alpha}{d^2},
\label{eq:iso_ratioFQ_FQsep_best_max}
\ee
which is $2$ for even $d$ and approaches $2$ for large $d$ for odd $d.$

{\it Proof.} Without the loss of generality, let us set $h_{\min}=-1$ and $h_{\max}=+1.$ Then, the denominator of \EQ{eq:RH2} is $8.$ Let us consider now the numerator. The maximum of the numerator of  \EQ{eq:RH2} will be clearly taken by a configuration for which $h_k=\pm 1.$ The first term is $d^2.$ Looking at the second term, we see that the numerator is maximized by $\{h_k\}_{k=1}^d=\{+1,-1,+1,-1,...\}.$ We find that the maximum is obtained for the Hamiltonian \eqref{eq:Hbest_anyd}.  $\qed$

Next, we determine which isotropic states are useful metrologically.

{\bf Observation S4.}---If
\be
p>p_{\rm m}=\frac{d^2-2}{4(d^2-\alpha)}+\sqrt{\frac{(d^2-2)^2}{16(d^2-\alpha)^2}+\frac{1}{d^2-\alpha}}
\label{eq:boundanyd_iso}
\ee holds
then the isotropic state $\varrho_{p}$ is useful for metrology with the Hamiltonian \eqref{eq:Hbest_anyd}. Otherwise, the isotropic state is not useful with any other Hamiltonian.

{\it Proof.} We look for the $p$ for which the righ-hand side of
\EQ{eq:iso_ratioFQ_FQsep_best_max} is $1.$ $\qed$

Note that $p_{\rm m}>1/2$ for all $d$ while for large $d$ it converges to $1/2.$ On the other hand, the isotropic state given in \EQ{eq:rhop} is entangled if $p>1/d.$ Hence, for all $d\ge 2$ there are isotropic states there are entangled but not useful for metrology.

Let us now look for the Hamiltonian of the type \eqref{eq:Asym} with which the isotropic states have the worst metrological performance.

{\bf Observation S5.}---Isotropic states have the worst metrological performance with respect to separable states with the Hamiltonian given by
\be
\mathcal H_{\rm worst}={\rm diag}(1,-1,0,0,...,0).
\label{eq:Hworst}
\ee
The corresponding metrological performance is described by 
\be
g(\varrho_p,\mathcal H^{({\rm iso})}_{\rm coll}({\mathcal H}_{\rm worst}))=\frac{4p^2d}{pd^2+2(1-p)}.
\label{eq:iso_ratioFQ_FQsep_worst_anyd}
\ee
No other Hamiltonian $\mathcal H$ corresponds to a worst performance.

Note that we considered collective Hamiltonians of the type \eqref{eq:Asym}. Other collective Hamiltonians $\mathcal H_{\rm coll}$  can lead to a worse performace and can even have $g(\varrho_p,\mathcal H_{\rm coll})=0.$ In particular, this is the case for Hamiltonians given in \EQ{eq:Aantisym}, where $\mathcal K$ can be any Hamiltonian.

The metrological gain given in \EQ{eq:iso_ratioFQ_FQsep_worst_anyd} is maximal for $p=1$ and has the value
\be
g(\varrho_p,{\mathcal H}^{({\rm iso})}_{\rm coll}({\mathcal H}_{\rm worst}))=\frac{4}{d}.\label{eq:gvarrho}
\ee
If $d\ge 4,$ then the right-hand side of \EQ{eq:gvarrho} is not larger than one. Hence, with $\mathcal H_{\rm worst},$ no isotropic state can be useful for $d\ge 4.$ For $d=3,$ on the other hand the right-hand side of \EQ{eq:gvarrho} is larger than one. Hence, for  $d=3,$ the maximally entangled state $\ket{\Psi^{({\rm me})}}$ is useful with the Hamiltonian $\mathcal H_{\rm worst}.$ We can also see that for $d=3$ the maximally entangled state $\ket{\Psi^{({\rm me})}}$ is useful with any Hamiltonian $\mathcal H_{\rm coll}^{({\rm iso})}.$

\begin{figure}
\includegraphics[width=7cm]{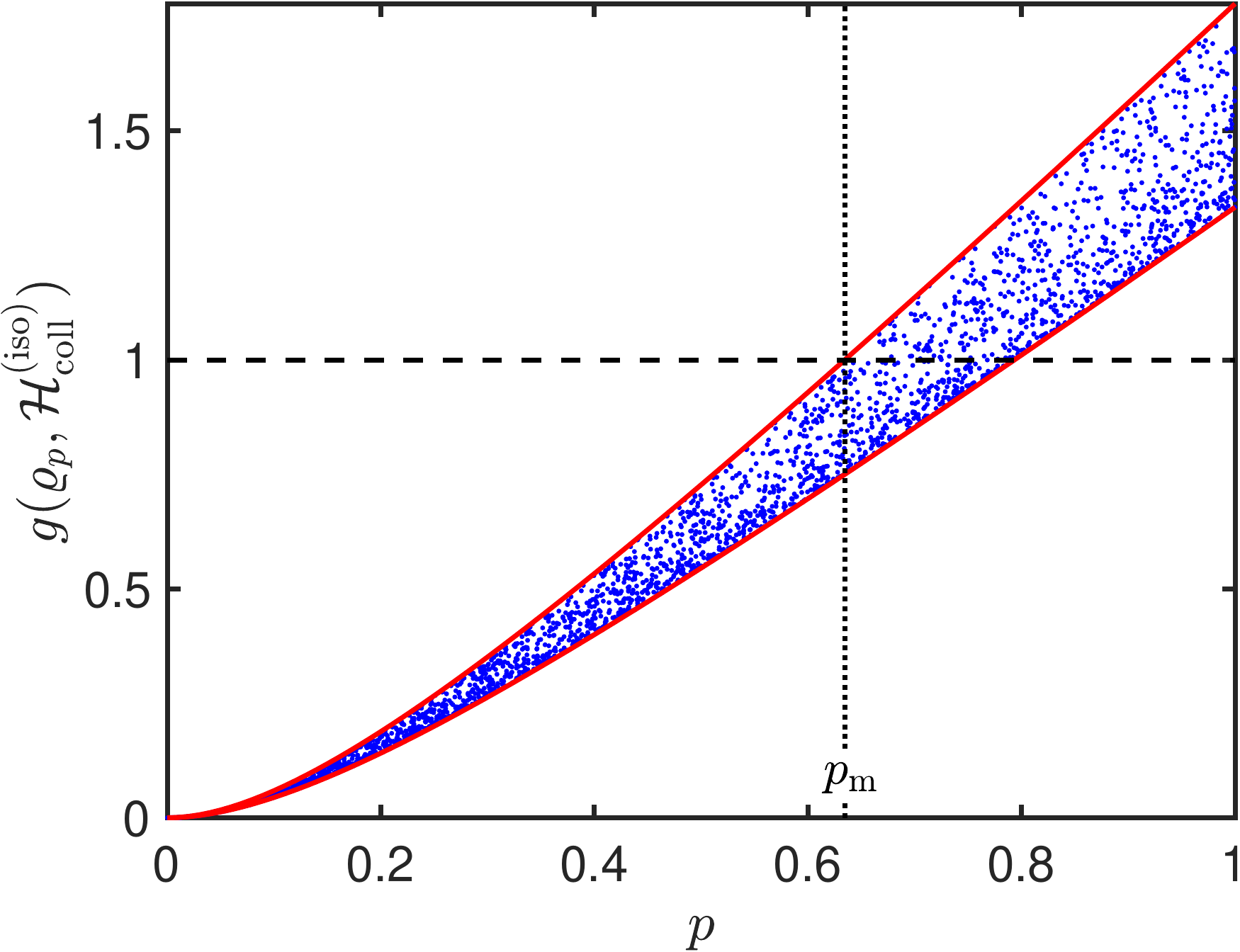}
\vskip0.5cm
\includegraphics[width=7cm]{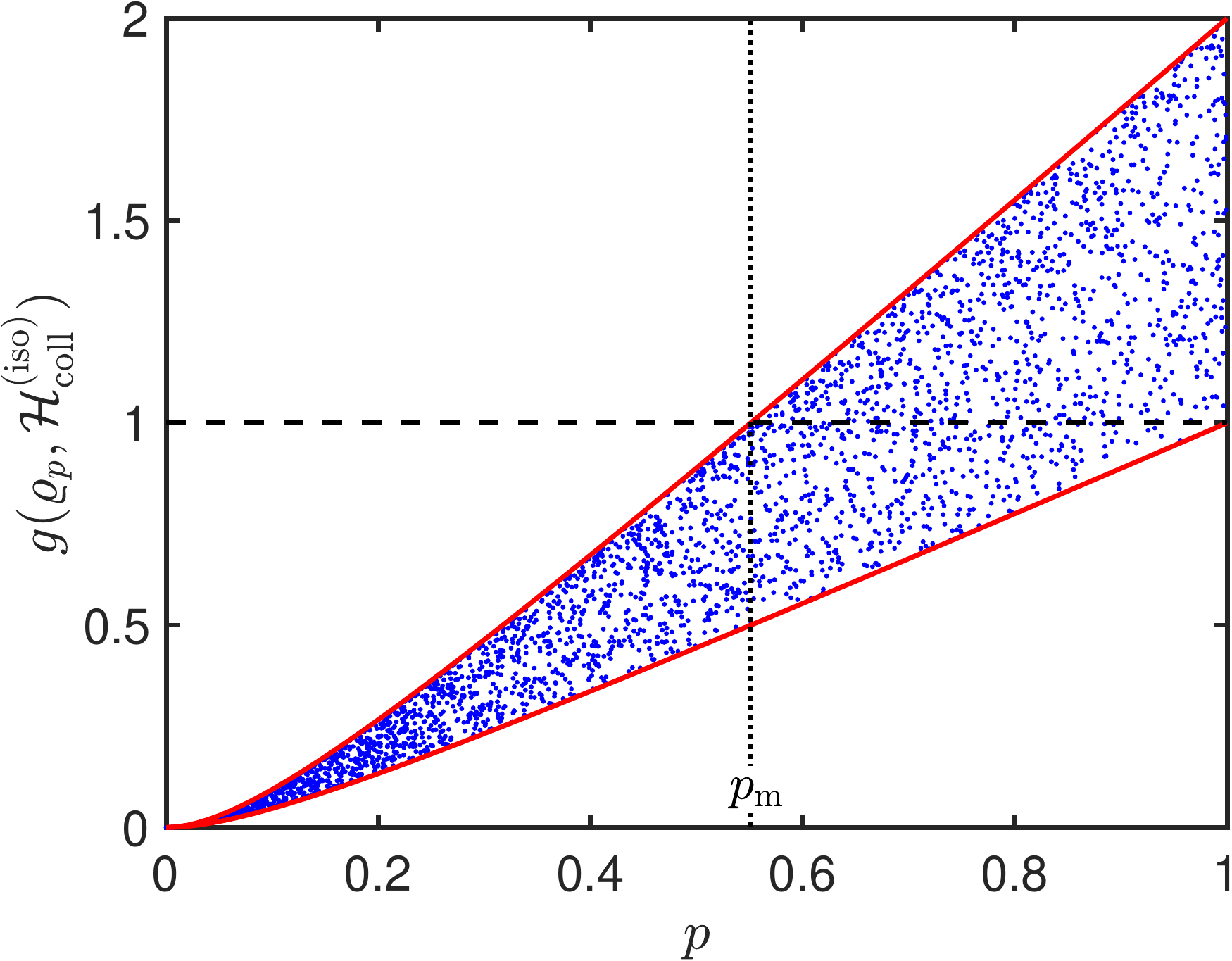}
\vskip0.5cm
\includegraphics[width=7cm]{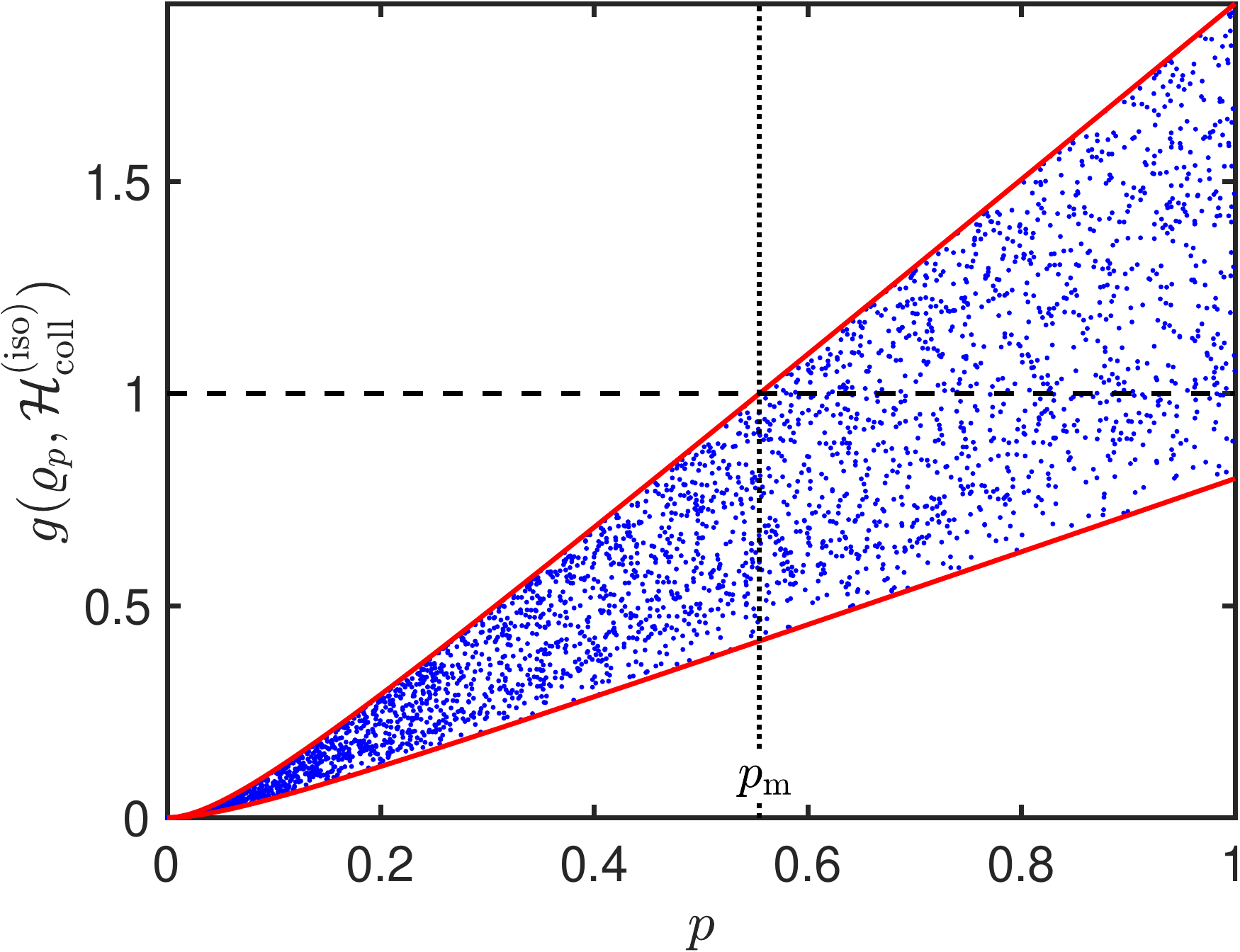}
\caption{Metrology with isotropic states given in \EQ{eq:rhop} for systems of size (top) $3\times 3$,
(middle) $4\times 4$, and
(bottom) $5\times 5.$
The metrological gain $g(\varrho_p,\mathcal H^{({\rm iso})}_{\rm coll})$ is plotted for isotropic states, \eqref{eq:rhop}, of a given $p.$ (dashed) Limit for separable states. (blue dots) Metrological performance of isotropic
states for two-body Hamiltonians $\mathcal H_{\rm coll}^{({\rm iso})}(\mathcal{H})$ given in \EQ{eq:Asym}, where
$\mathcal H$ are chosen randomly. (upper solid red line) Metrology with the best Hamiltonian $\mathcal H_{\rm best}$ given in \EQ{eq:Hbest_anyd}.
(lower solid red line) Metrology with the worst Hamiltonian $\mathcal H_{\rm worst}$ given in \EQ{eq:Hworst}. (dotted) Line corresponding the bound $p_{\rm m}$ given in \EQ{eq:boundanyd_iso}. Isotropic states with a larger $p$ are useful for metrology.}
\label{fig:iso}
\end{figure}

In \FIG{fig:iso}, we plot the results of simple numerics for $d=3,4$ and $5.$ The random mixed states have been generated according to \REF{Sommers2004Statistical}.

\section{Metrology with Werner states}
\label{app:met_W}

We now examine whether another type of bipartite states with a rotational symmetry, i.e, Werner states defined as \cite{Werner1989Quantum}
\be
\varrho_{\rm W}(\phi)=\frac{\eins+\phi V}{d^2+\phi d},\label{eq:Wernerstate}
\ee
outperform separable states in metrology.  Here $-1\le \phi \le +1$ and $V$ is the flip operator. 

We will consider a general evolution of the type \EQ{eq:Ageneral}.
Werner states are invariant under transformations of the type
\be
U\otimes U,
\label{eq:UU}
\ee where $U$ is a single-qudit unitary. Hence, Werner states are invariant under the Hamiltonian
\be
\mathcal H_{\rm inv}^{({\rm W})}(\mathcal{H})=\mathcal J  \otimes \mathbbm{1}+\mathbbm{1} \otimes \mathcal J,
\ee
where $\mathcal J$ is a Hermitian operator.

{\bf Observation S6.}---For short times, the action of the Hamiltonian $\mathcal{H}_{\rm coll}$ given in \EQ{eq:Ageneral}
is the same as the action of
\be
\mathcal H_{\rm coll}^{({\rm W})}(\mathcal{H})=\mathcal{H}^{({\rm W})} \otimes \mathbbm{1}-\mathbbm{1} \otimes \mathcal{H}^{({\rm W})},\label{eq:Aantisym2}
\ee
where the single party Hamiltonian $\mathcal H$ is defined as
\be
\mathcal{H}^{({\rm W})}=(\mathcal{H}_1+\mathcal{H}_2)/2. \label{eq:Hav2}
\ee

{\it Proof.} Let us define $\Delta^{({\rm W})}$ as
\be
\Delta^{({\rm W})} =(\mathcal H_2-\mathcal H_1)/{2}.\label{eq:Delta2}
\ee
In the rest of the section, we omit the superscript "W" in $\mathcal H_{\rm inv}^{({\rm W})}, \mathcal{H}^{({\rm W})}, \Delta^{({\rm W})}.$
Then, simple algebra shows that
\be
\mathcal{H}_{\rm coll}+\mathcal H_{\rm inv}^{({\rm W})}\left(\Delta^{({\rm W})} \right)=\mathcal H_{\rm coll}^{({\rm W})}.
\ee
Hence, for small $t$
\be
e^{-i \mathcal{H}_{\rm coll}t}e^{-i \mathcal H_{\rm inv} (\Delta) t}\approx e^{-i \mathcal H_{\rm coll}^{({\rm W})}(\mathcal{H})t}
\ee
holds. The Werner state is invariant under the action of $\mathcal H_{\rm inv}^{({\rm W})} (\Delta),$
since the corresponding unitary is of the form given in \EQ{eq:UU}.
Hence, the action of $\mathcal{H}_{\rm coll}$ is the same as the action of $H_{\rm coll}^{({\rm W})}(\mathcal{H})$ for small $t$.  $\qed.$

{\bf Observation S7.}---Replacing the evolution by $\mathcal{H}_{\rm coll}$ given in \EQ{eq:Ageneral} by the evolution by $\mathcal{H}_{\rm coll}^{({\rm W})}$ given in \EQ{eq:Aantisym} does not decrease the metrological gain. Hence, when looking for the Hamiltonian with the largest metrological gain, it is sufficient to look for Hamiltonians of the form \eqref{eq:Aantisym}.

{\it Proof.} The proof is similar to the proof of Observation S2. $\qed$

Werner states, given in  \EQ{eq:Wernerstate}, can also be defined as
\be\label{eq:Werner:PsPa}
\varrho_{\rm W}(\phi)=\frac{1+\phi}{d^2+\phi d} P_{\rm s} + \frac{1-\phi}{d^2+\phi d} P_{a},
\ee
where $P_{\rm s}$ and $P_{a}$ are the projectors to the symmetric and antisymmetric subspace, respectively. We will be interested in the case $\phi\le 0.$ The quantum Fisher information for Werner states for a Hermitian operator $A$ is
\begin{eqnarray}
F_{Q}[\varrho_{\rm W},A]&=&2\frac{(\lambda_{\rm s}-\lambda_{\rm as})^{2}}{\lambda_{\rm s}+\lambda_{\rm as}} \nonumber\\ & \times&\left(\sum_{k\in \mathcal S,l\in \mathcal A}\vert \langle k \vert A \vert l \rangle \vert^{2}+\sum_{k\in \mathcal A,l\in \mathcal S}\vert \langle k \vert A \vert l \rangle \vert^{2}\right),\nonumber\\
\end{eqnarray}
where $k\in \mathcal S$ and $l \in \mathcal A$ denote the indices of symmetric and antisymmetric eigenstates, respectively. From \EQ{eq:Werner:PsPa}, the eigenvalues of the Werner states can be obtained, yielding
\be
2\frac{(\lambda_{\rm s}-\lambda_{\rm as})^{2}}{\lambda_{\rm s}+\lambda_{\rm as}}
=\frac{4|\phi|^2}{d^2+\phi d}.
\ee

If the operator $A$ is of the form given in \EQ{eq:Aantisym}, then for any symmetric states $\ket{\Psi_{\rm s}}$ and antisymmetric states $\ket{\Psi_{a}}$
\be
\bra{\Psi_{\rm s}} A \ket{\Psi_{\rm s}}=\bra{\Psi_{a}} A \ket{\Psi_{a}}=0
\ee
hold.
Hence, we can return to sums over all eigenvectors and write
\begin{eqnarray}
F_{Q}[\varrho_{\rm W},\mathcal H_{\rm coll}^{({\rm W})}]&=&\frac{4|\phi|^2}{d^2+\phi d} \sum_{k,l}\vert \langle k \vert \mathcal H_{\rm coll}^{({\rm W})}\vert l \rangle \vert^{2}\nonumber\\
&=&\frac{8|\phi|^2}{d^2+\phi d}\trace(\mathcal (H_{\rm coll}^{({\rm W})})^2).
\end{eqnarray}
Then, we need that
\begin{eqnarray}
\trace((\mathcal H_{\rm coll}^{({\rm W})})^2)
&=&2[d \trace(\mathcal H^2)-\trace(\mathcal H)^2].
\end{eqnarray}
Hence, we obtain a general formula for the quantum Fisher information for Werner states as
\begin{equation}
F_{Q}[\varrho_{\rm W},\mathcal H_{\rm coll}^{({\rm W})}(\mathcal{H})]=\frac{8|\phi|^2}{d^2+\phi d}[d \trace(\mathcal H^2)-\trace(\mathcal H)^2].\label{eq:FQH}
\end{equation}

Based on \EQ{eq:FQH} and on \EQ{eq:seplim}, the metrological performance is given by
\be
g(\varrho_{\rm W},\mathcal H_{\rm coll}^{({\rm W})}(\mathcal H))=\frac{8|\phi|^2}{d^2+\phi d} r(\mathcal H),\label{eq:performance}
\ee
where $r(\mathcal H)$ is defined in \EQ{eq:RH2}.

Let us now look for the Hamiltonian that provides the largest metrological gain for Werner states.

\begin{figure}
\includegraphics[width=7cm]{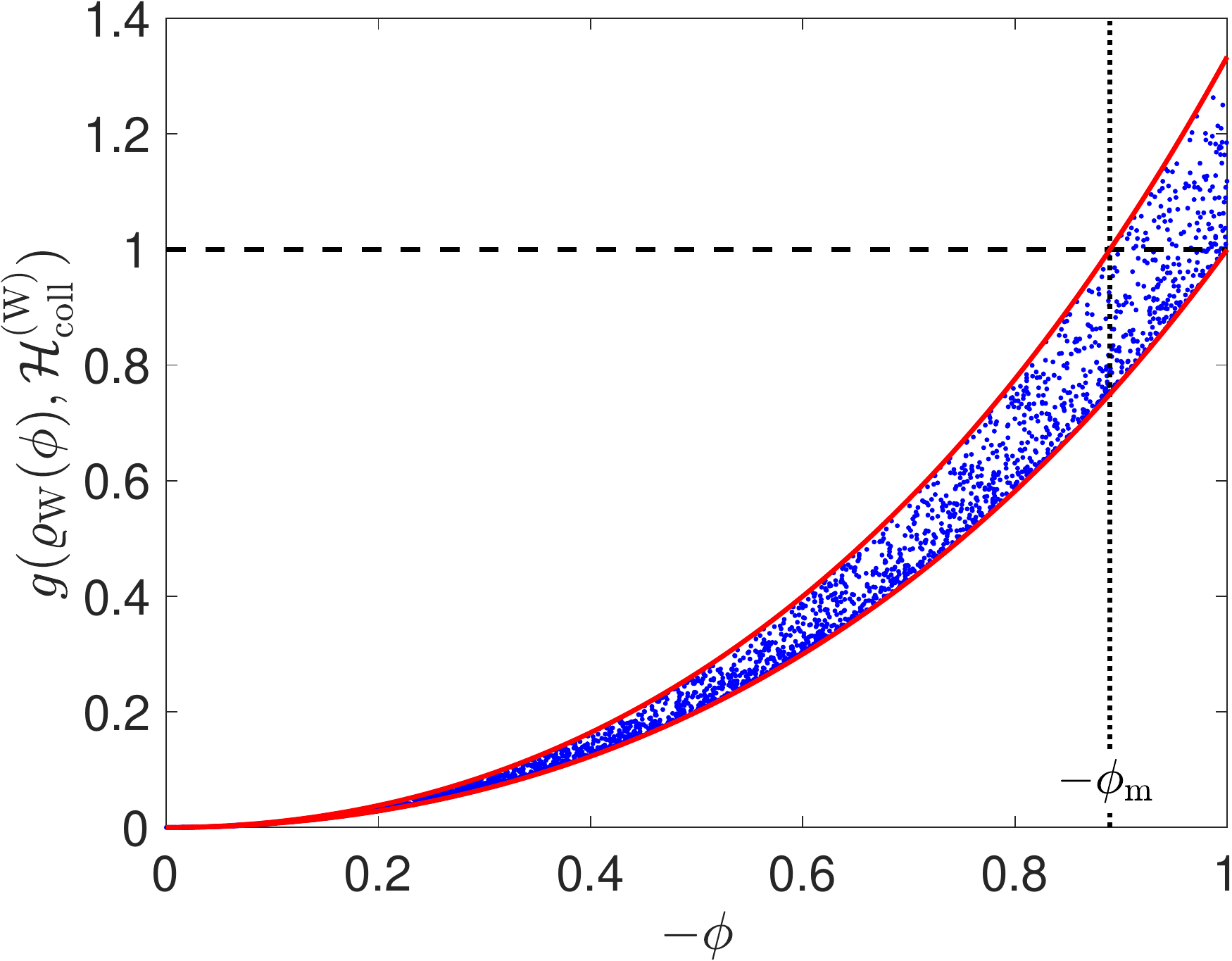}
\vskip0.5cm
\includegraphics[width=7cm]{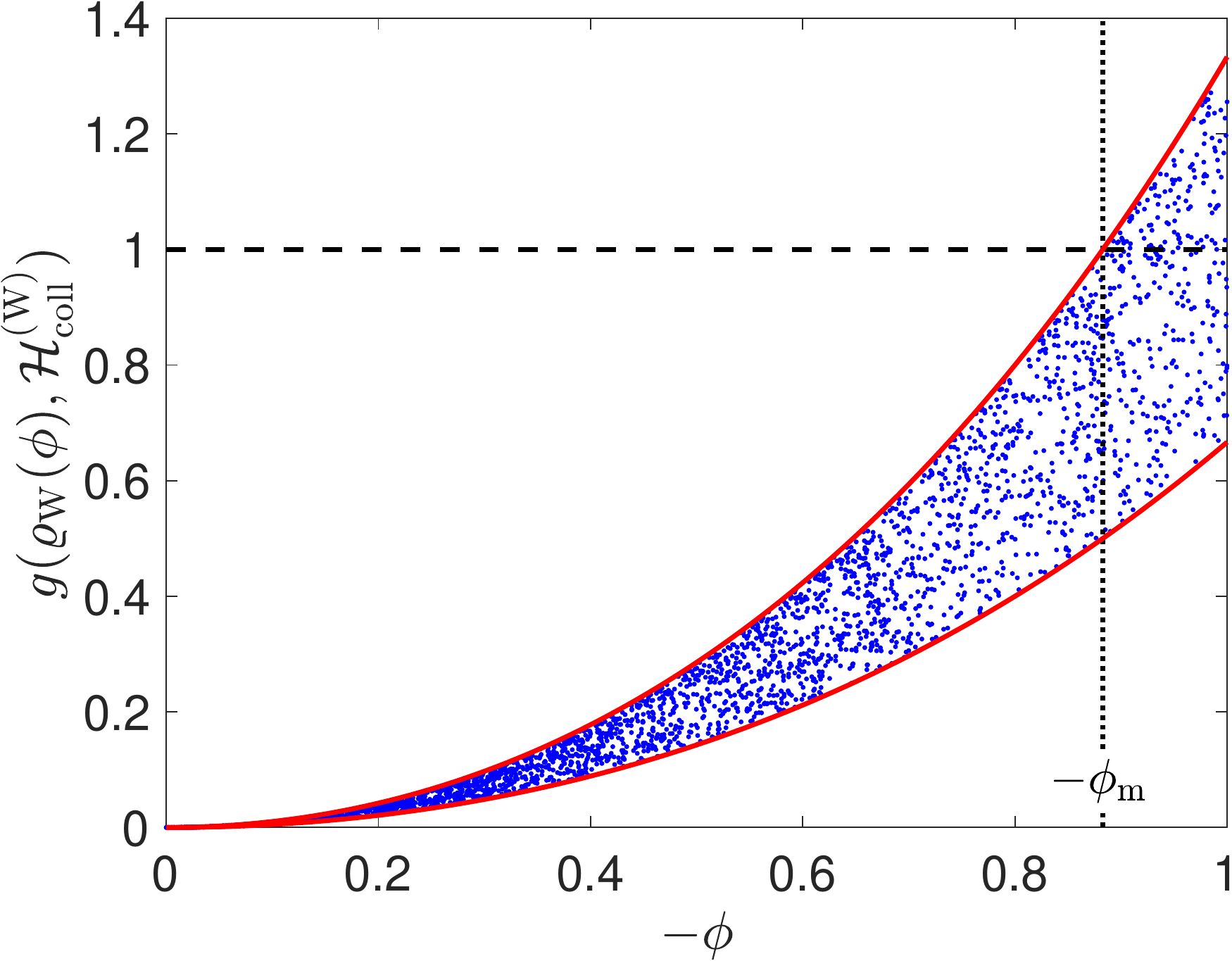}
\vskip0.5cm
\includegraphics[width=7cm]{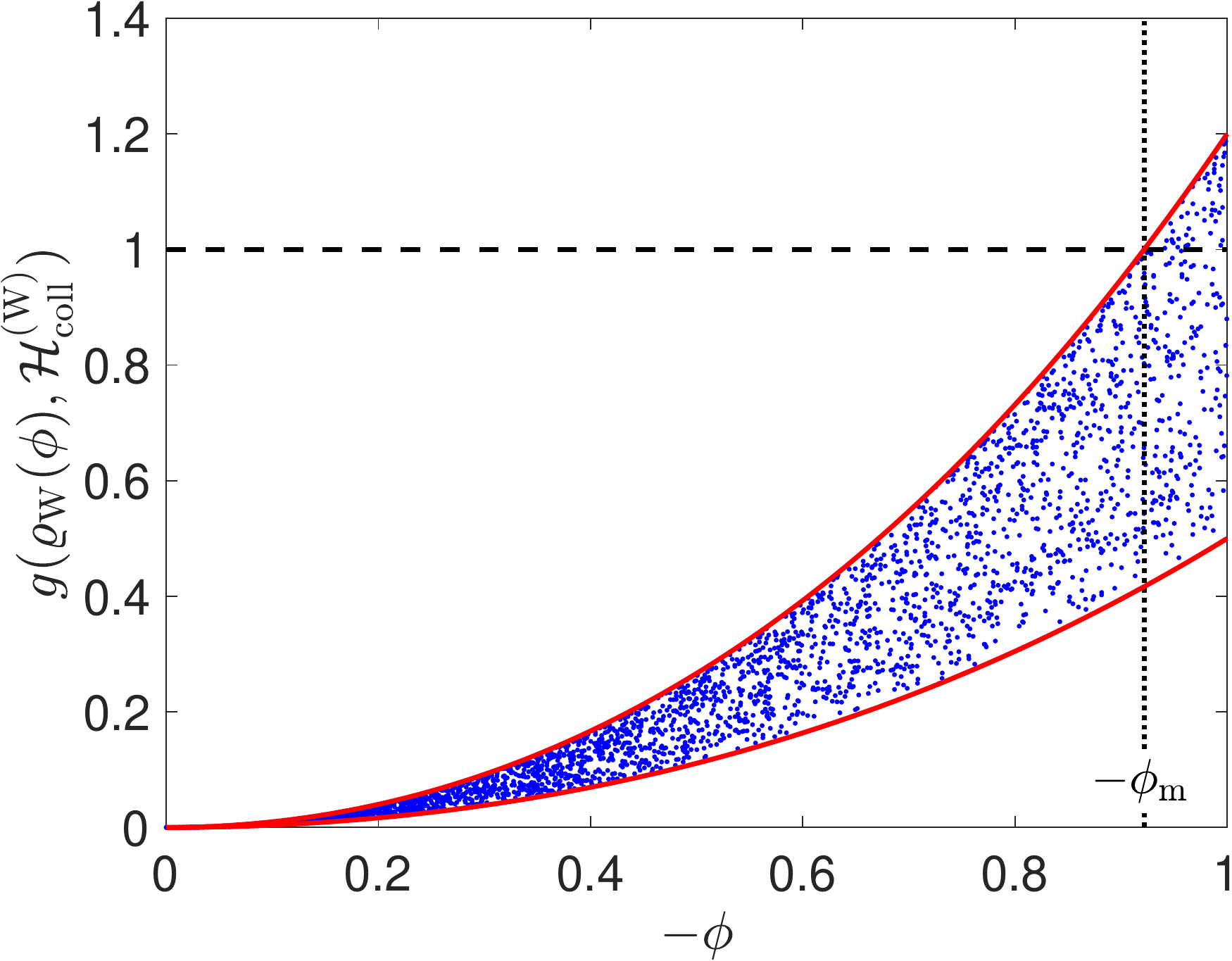}
\caption{Metrology with Werner states given in \EQ{eq:Wernerstate}. (top) $3\times 3$,
(middle) $4\times 4$, and
(bottom) $5\times 5$ Werner states are considered.
The metrological gain $g(\varrho_{\rm W}(\phi),\mathcal H^{({\rm W})}_{\rm coll})$ is plotted for Werner states of a given $\phi.$ (dashed) Limit for separable states. (blue dots)
Metrological performance of Werner
states for two-body Hamiltonians $\mathcal H_{\rm coll}^{({\rm W})}(\mathcal{H})$ given in \EQ{eq:Aantisym}, where
$\mathcal H$ are chosen randomly.
 (upper solid red line) Metrology with the best Hamiltonian $\mathcal H_{\rm best}$ given in \EQ{eq:Hbest_anyd}.
(lower solid red line) Metrology with the worst Hamiltonian $\mathcal H_{\rm worst}$ given in \EQ{eq:Hworst}. (dotted) Line corresponding the bound $\phi_{\rm m}$ given in \EQ{eq:boundanyd}. Werner states with $-\phi>-\phi_{\rm m}$ are useful for metrology.}
\label{fig:Werner}
\end{figure}

{\bf Observation S8.}---Werner states have the best metrological performance with respect to separable states with the Hamiltonian $\mathcal H_{\rm best}$ given in \EQ{eq:Hbest_anyd}.
The corresponding quantum Fisher information is
\be
g(\varrho_{\rm W},\mathcal H_{\rm coll}^{({\rm W})}(\mathcal{H}_{\rm best}))=\frac{|\phi|^2(d^2-\alpha)}{d^2+\phi d},
\label{eq:ratioFQ_FQsep_best_evend}
\ee
where the optimization is carried out over collective Hamiltonians of the form  \eqref{eq:Aantisym}.

No other such collective Hamiltonian corresponds to a better performance. \EQL{eq:ratioFQ_FQsep_best_evend} is maximal for $\phi=-1$ and has the value
\be
g(\varrho_{\rm W},\mathcal H_{\rm coll}^{({\rm W})}(\mathcal{H}_{\rm best}))=\frac{d+\alpha}{d+\alpha-1},
\ee
which is close to $1$ for large $d.$

{\it Proof.} The best $\mathcal H$ operator is the one for which $r(\mathcal H)$ defined in \EQ{eq:RH2}
is the largest. In other words, we can look for the $\mathcal H$ for a constant $(h_{\max}-h_{\min})^2$ that maximizes $[d \trace(\mathcal H^2)-\trace(\mathcal H)^2].$
The details of the proof are similar to the proof of Observation S3.
$\qed$

Next, we determine which Werner states are useful metrologically.

{\bf Observation S9.}---If
\be
\phi<\phi_{\rm m}:=
\frac{d}{2(d^2-\alpha)} - \sqrt{\frac{d^2}{4(d^2-\alpha)^2}+\frac{d^2}{d^2-\alpha}}\label{eq:boundanyd}
\ee
holds,
then the Werner state is useful for metrology with the Hamiltonian \eqref{eq:Hbest_anyd}.
Otherwise, the Werner state is not useful with any other Hamiltonian.

{\it Proof.} We look for the $\phi$ for which the right-hand side of
\EQ{eq:ratioFQ_FQsep_best_evend} is $1.$  $\qed$

Let us now look for the Hamiltonian of the type \eqref{eq:Aantisym} with which the Werner states have the worst metrological performance.

Note that for large $d$ the parameter $\phi_{\rm m}$ converges to $1,$ while Werner states are entangled if $\phi<-1/d$ \cite{Werner1989Quantum}. Hence, there are Werner states that are entangled but not useful for metrology.

{\bf Observation S10.}---Werner states have the worst metrological performance with respect to separable states with the Hamiltonian given in \EQ{eq:Hworst}.
The corresponding metrological gain is
\be
g(\varrho_{\rm W},\mathcal H_{\rm coll}^{({\rm W})}(\mathcal{H}_{\rm worst}))=\frac{2|\phi|^2d}{d^2+\phi d}.
\label{eq:ratioFQ_FQsep_worst_anyd}
\ee
No other Hamiltonian corresponds to a worst performance.

{\it Proof.} This can be seen noting that \EQ{eq:performance} is minimized for this case.  $\qed$

Note that we considered Hamiltonians $\mathcal H_{\rm coll}^{({\rm W})}(\mathcal{H})$ of the type
\eqref{eq:Aantisym}. Other collective Hamiltonians $\mathcal H_{\rm coll}$ can lead to a worse performace and can even reach to $g(\varrho_{\rm W},\mathcal H_{\rm coll})=0.$
In particular, this is the case for collective Hamiltonian of the form given in \EQ{eq:Aantisym}.

\EQL{eq:ratioFQ_FQsep_worst_anyd} is maximal for $\phi=-1$ and has the value
\be
g(\varrho_{\rm W},\mathcal H_{\rm coll}^{({\rm W})}(\mathcal{H}_{\rm worst}))=\frac{2}{d-1}.\label{eq:limitworst}
\ee
We can see that for $d\ge 3$ the right-hand side of \EQ{eq:limitworst} is not larger than one,
hence the Werner state is not useful with the Hamiltonian $\mathcal{H}_{\rm worst}.$
We can also see that the metrological gain, \eqref{eq:limitworst}, is close to $0$ for large $d.$

In \FIG{fig:Werner}, we plot the results of simple numerics for $d=3,4$ and $5.$
The random mixed states have been generated according to \REF{Sommers2004Statistical}.

\section{Concrete example with two-qubit singlets}
\label{app:2ubit}

In this Section, we work out in detail the problem of metrology with two-qubit singlets and ancillas.
This problem is also interesting, since the Hamiltonians obtained numerically
are very simple.

Let us consider the noisy two-qubit singlet
\begin{equation}
\varrho^{(p)}_{AB}=(1-p)\ketbra{\Psi^{-}}+p\openone/4, \label{eq:nsinglet}
\end{equation}
where
\begin{equation}
\ket{\Psi^{-}}=\frac 1{\sqrt{2}}(\ket{01}-\ket{10}).
\end{equation}
The state given in \EQ{eq:nsinglet} is a Werner-state given in \EQ{eq:Wernerstate} and it is also equivalent to 
an isotropic state, \eqref{eq:rhop}, under local unitaries. The state is more useful than separable states if 
the noise is smaller than
\be
p_{\rm limit}=\tfrac1 8(7-\sqrt{17})\approx 0.3596,\label{eq:plimit}
\ee
see \EQ{eq:boundanyd_iso} for isotropic states. The optimal local Hamiltonian is 
\be
\mathcal H_{\text{1\;singlet}}=Z_{a}-Z_{B},\label{eq:Hopt_1singlet}
\ee
where $Z$ is the Pauli spin matrix ${\rm diag}(-1,+1).$ Even for a pure singlet, this is the optimal Hamiltonian.

Let us consider two singlets with a bipartition $AA'|BB'$
\be
\varrho_{\text{2\;singlets}}=\varrho^{(p)}_{AB}\otimes \varrho^{(p)}_{A'B'}.
\ee
Then, the optimal Hamiltonian is
\be
\mathcal H_{\text{2\;singlets}}=Z_{a}Z_{A'}+Z_{B}Z_{B'}.
\ee
 
Finally, let us consider a singlet in $AB$ and two ancillas in some pure state in $A'B'$
\be
\varrho^{(p)}_{AB}\otimes \ketbra{\Psi_{A'}}\otimes \ketbra{\Psi_{B'}}.
\ee
In this case, if $p<p_{\rm limit}$ then the optimal Hamiltonian is \EQ{eq:Hopt_1singlet}. That is, the ancillas do not give any advantage, the Hamiltonian does not act on the ancillas. If the singlet is too noisy, that is, $p>p_{\rm limit}$ then the optimal local Hamiltonian is of the form
\be
\mathcal H_{A'}+\mathcal H_{B'}.\label{eq:toonoisy}
\ee
Note that \EQ{eq:toonoisy} acts only on the ancillas.

If we use pure singlets then in all these cases we have $\FQ=16,$ while the limit for separable states is $\FQ^{(\rm sep)}=8.$
If we use singlets with $p$ given in \EQ{eq:plimit}, then
\be
\FQ[\varrho_{\text{2\;singlets}},\mathcal H_{\text{2\;singlets}}]=8.1530.
\ee
Thus, the state outperfoms separable states.
In the case of a single copy, and a single copy with two pure ancillas, 
$\FQ=\FQ^{(\rm sep)}=8.$
On the other hand, the state $\varrho_{\text{2\;singlets}}$ remains more useful than separable states if
\be
p<0.3675,
\ee
where the limit on the noise fraction has been obtained numerically.

Thus, in the $2\times2$ case, a singlet mixed with white noise cannot be activated by ancillas. This is also true for isotropic states, since they are locally equivalent to a singlet mixed with white noise.  

Finally, we show that if a singlet is mixed with non-white noise, then it can be activated with ancillas. Let us consider the state
\be
 \frac{1}{2}\left(\ketbra{\Psi^-}+\ketbra{00}\right). \label{eq:nonsymnoise}
\ee
For this state, the optimization over Hamiltonians lead to $\FQ = 8,$ which is also the bound for separable states, i.e.,  $\FQ^{(\rm sep)}=8.$ With two ancillas we can reach $\FQ = 9.$ With two copies of the state \EQ{eq:nonsymnoise}, we can reach $\FQ= 10.$ In all these cases, we could use $c_1=c_2=1$ when searching for the optimal Hamiltonian due to the symmetries of the setup. [See \EQ{eq:cnHconst} for the definition of $c_k.$] The state given in \EQ{eq:nonsymnoise} can be activated even with a single ancilla.  By setting $c_1=c_2=1,$ we get $\FQ = 8.4.$ On the other hand, the optimal Hamiltonian has $c_1=1$ and $c_2=(1+\sqrt5)/2 \approx 1.618$ and  and the gain reaches $\FQ/\FQ^{(\rm sep)}=3(5+\sqrt 5)/20\approx 1.0854.$

We considered various multiqubit states in this section. In an application, we have to choose one of them. The basic idea is the following. If the metrological gain of an entangled quantum state is not larger than 1, i.e., $g\le 1,$ then it is better to use product states since they can reach the same precision, but it is easier to create them. Moreover, if we find that an entangled state is more useful than separable states,  i.e., $g> 1,$ then our algorithm can also tell us the optimal Hamiltonian corresponding to the task where they outperform separable states the most.

\section{Estimation of the metrological gain for general quantum states}
\label{app:Estimation of the quantum Fisher information for general quantum states}

Recently, there have been several methods presented to find lower bounds on the quantum Fisher information based on few operator expectation values \cite{Pezze2009Entanglement,Apellaniz2017Optimal}. Our results on isotropic states and Werner states  can be used to construct lower bounds for the metrological gain $g$ based on a single operator expectation value.

In order to proceed, we note that any $d\times d$ state can be depolarized into an isotropic state given in \EQ{eq:rhop} with the $U\otimes U^*$ twirling operation as
\be 
\label{twirling}
\varrho_{{\rm iso}}(F)=\int \mathcal M(dU) (U\otimes U^*) \varrho (U^{\dagger}\otimes U^{*\dagger}),
\ee 
where $\mathcal M$ is a unitarily invariant probability measure. 
The state $\varrho_{{\rm iso}}(F)$ is just the isotropic state given in \EQ{eq:rhop}, defined with a different parametrization as
\be\label{isotrop} \varrho_{{\rm iso}}(F)=F\ket{\Psi^{({\rm me})}}\bra{\Psi^{({\rm me})}}+(1-F)\frac{\openone-\ket{\Psi^{({\rm me})}}\bra{\Psi^{({\rm me})}}}{d^2-1},
\ee 
where the maximally entangled state $\ket{\Psi^{({\rm me})}}$ is given in \EQ{eq:mestate},
and 
\be 
F={\rm Tr}(\varrho \ketbra{\Psi^{({\rm me})}}) \label{eq:Fproj}
\ee
is the entanglement fraction of the state $\varrho$, which is alternatively called the singlet fraction \cite{Horodecki1999Reduction,Horodecki1999General}.  Based on \EQ{eq:iso_ratioFQ_FQsep_best}, the maximum metrological performance of the isotropic state is given by
\be
\label{giso}
g(\varrho_{{\rm iso}}(F)) = \frac{2(d^2-\alpha)(d^2F-1)^2}{d^2(d^2-1)(1-2F+d^2F)},
\ee
where $\alpha$ is zero for even $d$, and one otherwise. Here, we remember that the metrological gain is defined in \EQ{eq:glocalH}.

Next, we show that $g(\varrho)$ cannot increase under twirling defined in \EQ{twirling}, i.e.,
\be
\label{gg}
g(\varrho)\ge g(\varrho_{\rm iso}(F)).
\ee
We use a series of inequalities
\begin{align}
\FQ[\rho_{p},\mathcal H]
&=\FQ\left[\int \mathcal M(dU) (U\otimes U^*) \varrho (U^{\dagger}\otimes U^{*\dagger}),\mathcal H\right]\nonumber\\
&\le\int  M(dU) {\FQ[(U\otimes U^*) \varrho (U^{\dagger}\otimes U^{*\dagger}),\mathcal H]}\nonumber\\
&\le{\FQ[(U_0\otimes U_0^*) \varrho (U_0^{\dagger}\otimes U_0^{*\dagger}),\mathcal H]}\nonumber\\
&=\FQ[\varrho,\mathcal H'],
\end{align}
where $\mathcal H'=(U_0^{\dagger}\otimes U_0^{*\dagger}) \mathcal H (U_0\otimes U_0^*)$ and $U_0$ is some unitary.
To arrive at the second line we used the property of the quantum Fisher information that it is convex in the state, 
Noting also that the eigenvalues of $\mathcal H'$ are the same as that of $\mathcal H$, and that $\FQ^{({\rm sep})}(\mathcal H)$ in \EQ{eq:seplim} depends only on the eigenvalues, we arrive at \EQ{gg}.

Based on \EQ{gg}, the metrological gain of any quantum state can be bounded from below as
\be
g(\varrho) \ge g(\varrho_{{\rm iso}}(F)),
\ee
where $g(\varrho_{{\rm iso}}(F))$ is defined in \EQ{giso} and $F$ is just the entanglement fraction of $\varrho.$
Based on \EQ{eq:Fproj}, $F$ equals the expectation value of the projector to $\ket{\Psi^{({\rm me})}}.$ Hence, our lower bound is based on a single operator expectation value.

Similar calculations can be carried out for Werner states, using the fact that any quantum state can be depolarized into a Werner state using the $U\otimes U$ twirling
\be 
\label{twirling2}
\varrho_{{\rm W}}(\phi)=\int \mathcal M(dU) (U\otimes U) \varrho (U^{\dagger}\otimes U^{\dagger}).
\ee 
Then, we can construct a lower bound
\be
g(\varrho) \ge g(\varrho_{{\rm W}}(\phi)),\label{eq:boundwww}
\ee
where the \EQ{eq:ratioFQ_FQsep_best_evend} gives the right-hand side of \EQ{eq:boundwww} as a function of the parameter $\phi.$
The quantity $\phi$ is related to the expectation value of the flip operator $V$ as
\be
\exs{V}=\frac{1+d\phi}{d+\phi}.
\ee

\section{Uniting qudits}

In most of the paper, we considered bipartite examples. 
In the multipartite case, the usefulness of a quantum state is always relative to the partitioning of the parties. From this point of view, it is worth to look at metrological usefulness of a multipartite state when we put the parties into two groups, and return to the bipartite problem.
For instance, the four-qubit ring cluster state is not useful, $\FQ/\FQ^{({\rm sep})}=1$ \cite{Hyllus2010Not}. After uniting two qubits into a ququart it becomes useful, with $\FQ/\FQ^{({\rm sep})}=2.$
An optimal Hamiltonian with an optimal gain is
\be
j_z^{(1)}\otimes j_y^{(2)}+j_y^{(3)}\otimes j_z^{(4)}.\label{eq;Hxxxx}
\ee
We have to measure $M=j_z^{(1)}\otimes j_x^{(2)}\otimes j_x^{(3)}\otimes j_z^{(4)}$ for an optimal estimation precision $\va{\theta}_M=1/16.$
Due to the commutator relations $[j_z^{(n)},M]=[j_z^{(n)},\mathcal H]=0$ for $n=1,4$, we can realize the following scheme.
We measure $j_z$ on qubits (1) and (4) such that we have a state locally equivalent to a singlet on qubits (2) and (3). Then, we do metrology with qubits (2) and (3).
Similar schemes based on preselection have appeared in the theory of entanglement and nonlocality 
\cite{Toth2006Two-setting,Popescu1992Generic}.

\section{How large part of quantum states are useful}

\begin{figure}[t!]
\centerline{
\epsfxsize4cm \epsffile{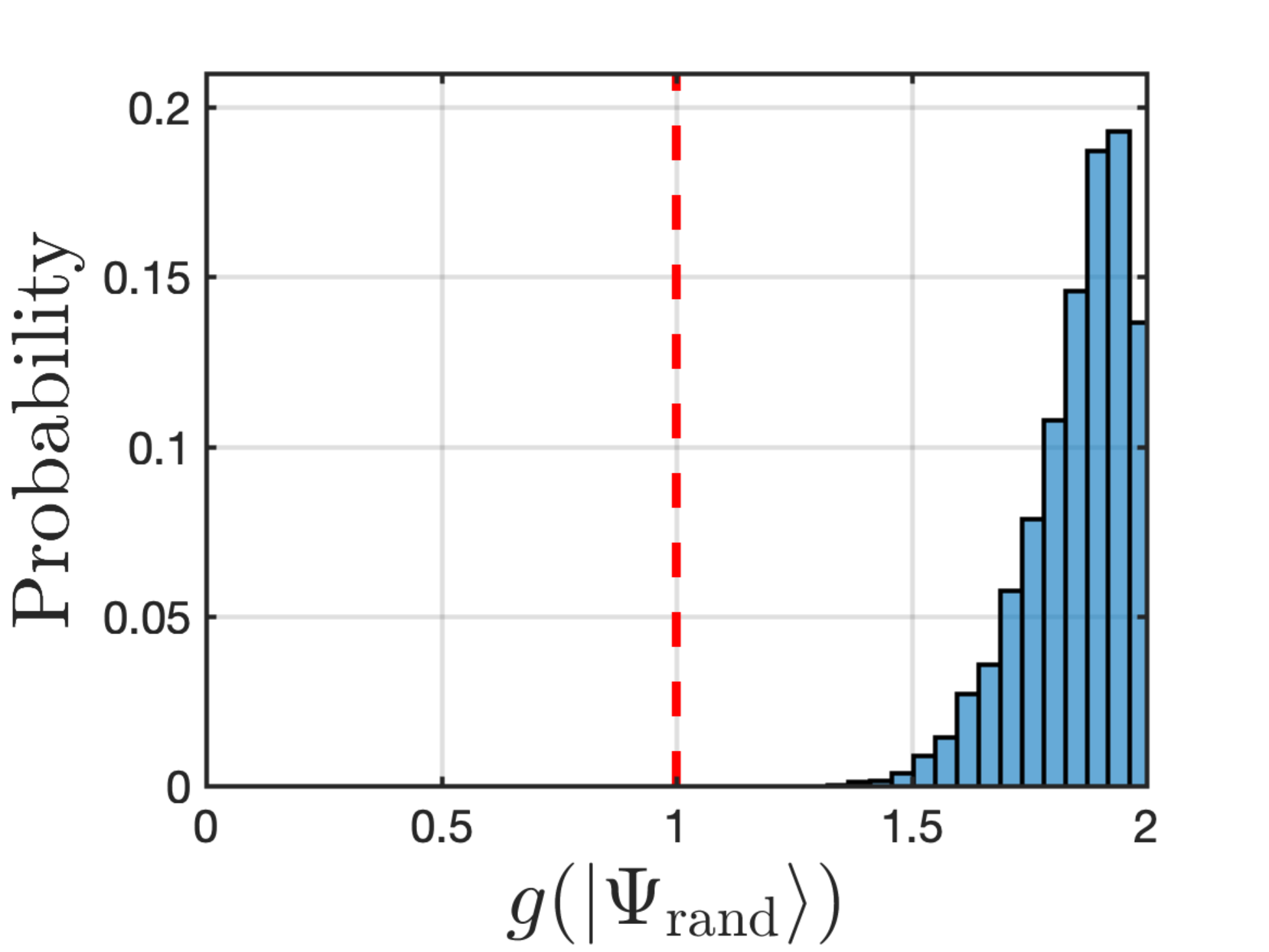}\hskip0.5cm
\epsfxsize4cm \epsffile{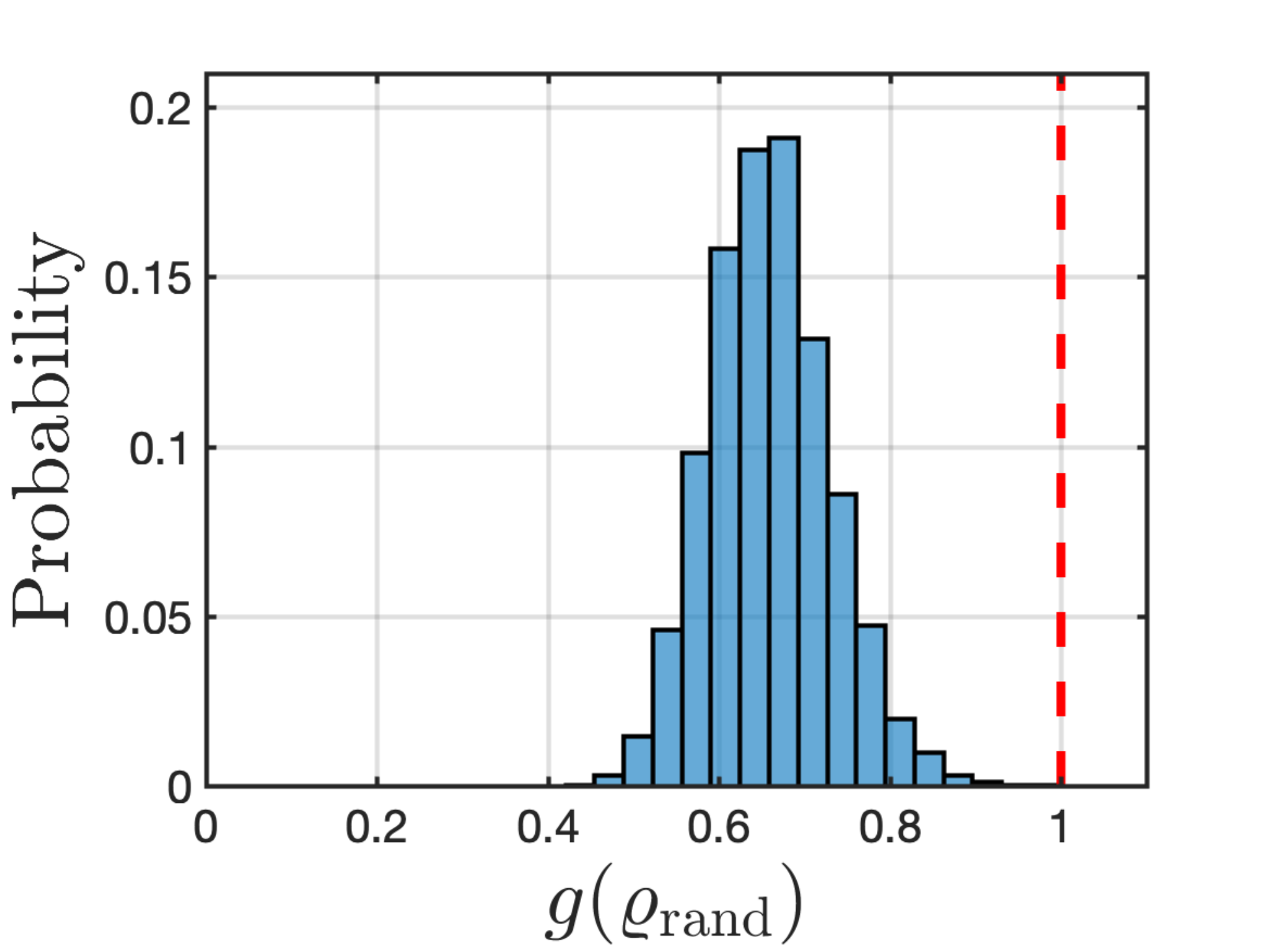}}
(a) \hskip4cm (b)
\centerline{
\epsfxsize4cm \epsffile{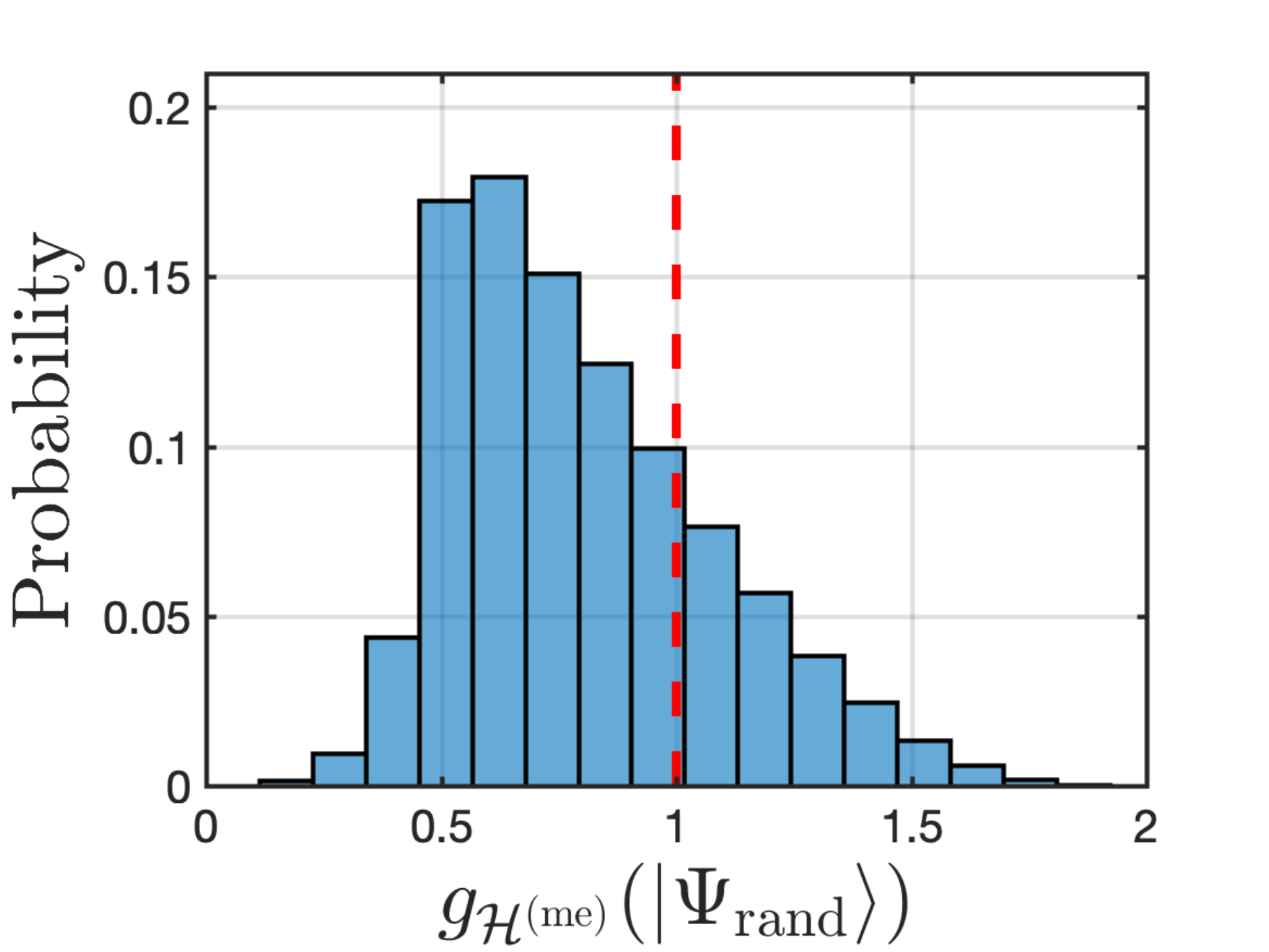}\hskip0.5cm
\epsfxsize4cm \epsffile{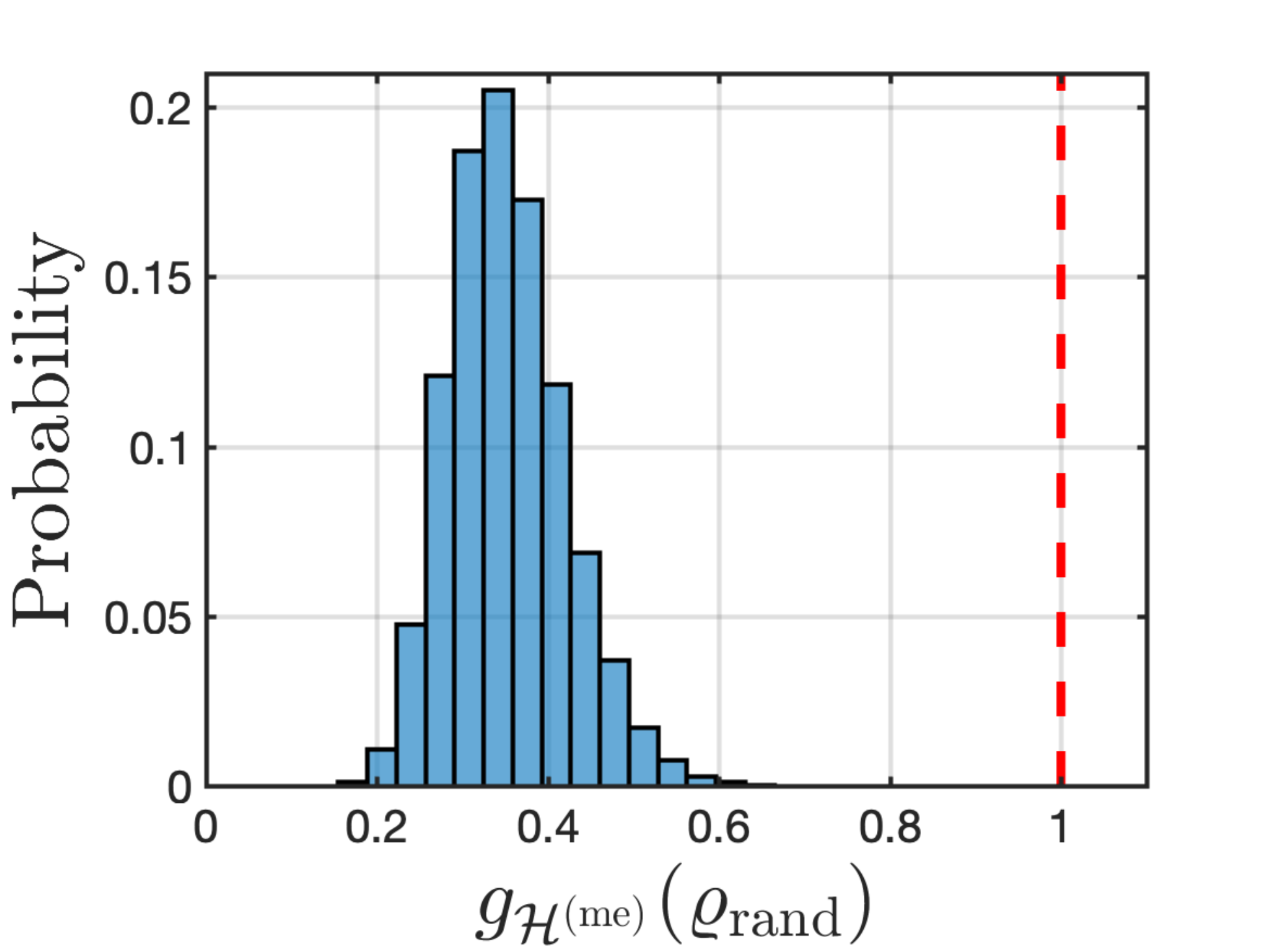}
}
(c) \hskip4cm (d)
\caption{Distribution of the metrological gain optimized over local Hamiltonians. Results for random states with dimension $3\times3$ for (a) pure and (b) mixed states.
(c) and (d) The same for the Hamiltonian given in \EQ{eq:Hme}. (dashed vertical line) Line corresponding to $g=1.$ States are metrologically useful if $g>1.$ 
} \label{fig:statistics}
\end{figure}

The scaling of the quantum Fisher information with the dimension has been considered for random states and for the best local Hamiltonian in  \REF{Oszmaniec2016Random}.
We used our optimization algorithm to determine the distribution of the  quantum Fisher information and obtain exactly how large part of pure or mixed quantum states are useful. The random pure states and mixed states have been generated according to \REF{Sommers2004Statistical}.
For $d=3,$ the results are shown in
 \FIG{fig:statistics}.  It suggests that almost no random mixed states are useful.
Pure states are useful almost with a maximal usefulness.

\section{Infinite number of copies of arbitrary pure states}
\label{app:Infinite number of copies of arbitrary pure states}

It is shown that an infinite number of copies of any entangled pure quantum state of Schmidt rank-$s$ with $s>1$ is maximally useful metrologically. To this end, let us define a pure state in the Schmidt basis with Schmidt rank-$s$ as in \EQ{schmidt}. Here, the real non-negative $\sigma_k$ Schmidt coefficients are in a descending order, and $\sum_{k=1}^s\sigma_k^2=1$. In addition, we also assume that $\sigma_1>\sigma_2$.

Then, the $n$-copy state has the Schmidt coefficients
\be
\sigma_{i_1}\sigma_{i_2}\cdot\cdot\cdot\sigma_{i_n},
\ee
where $i_k\in\{1,2,\ldots,s\}.$
The number of equal Schmidt coefficients in the $n$-copy state follows a multinomial distribution formula. With this and
\EQ{evenS}, we obtain the lower bound
\begin{align}
&\FQ[\ket{\psi}^{\otimes n},\mathcal H_{n-{\rm copy}}] \ge\nonumber\\
&8\sum_{k_1+k_2+\ldots+k_s=n}^n\left\lfloor\frac{1}{2}\binom{n}{k_1,k_2,\ldots,k_s}\right\rfloor(2\sigma_1^{k_1}\sigma_2^{k_2}\cdot\cdot\cdot\sigma_s^{k_s})^2,
\label{eq:fqncopy}
\end{align}
where 
\begin{eqnarray}\label{eq:HABncopy}
&&\mathcal H_{n-{\rm copy}}=\left(\bigotimes_{k=1}^n \mathcal H_{{\rm A},k}\right)\otimes
\left(\bigotimes_{k=1}^n \openone_{{\rm B},k}\right)+\nonumber\\
&&\left(\bigotimes_{k=1}^n \openone_{{\rm A},k}\right)\otimes
\left(\bigotimes_{k=1}^n \mathcal H_{{\rm B},k}\right).
\end{eqnarray}
Here $\mathcal H_{{\rm A},k}=\mathcal H_{{\rm B},k}$ are all equal to the operator given in 
in \EQ{eq:HA}. $\mathcal H_{{\rm A},k}$ and $\mathcal H_{{\rm B},k}$ act on the $kth$ copy of system, on  subsystem $A$ and $B,$ respectively.
The meaning of $\openone_{{\rm A},k}$ and $\openone_{{\rm B},k}$ is analogous.
The expression $\left\lfloor x \right\rfloor$ is the floor or integer part of $x$, and the multinomial coefficients are
\begin{align}
\binom{n}{k_1,k_2,\ldots,k_s}=\frac{n!}{k_1!k_2!\cdot\cdot\cdot k_s!}.\label{eq:multinom}
\end{align}

Using the multinomial theorem for $\left(\sum_k{\sigma_k^2}\right)^{n}=1$ and the relation
\be
\left\lfloor\frac{1}{2}\binom{n}{k}\right\rfloor\ge \frac{\binom{n}{k}-1}{2}, 
\ee
yield a further lower bound
\begin{align}
&\FQ[\ket{\psi}^{\otimes n}, \mathcal H_{n-{\rm copy}}]\nonumber\\
&\ge16\sum_{k_1+k_2+\ldots+k_s=n}{\left[\binom{n}{k_1,k_2,\ldots,k_s}-1\right]}{\sigma_1^{2k_1}\sigma_2^{2k_2}\cdot\cdot\cdot\sigma_s^{2k_s}}\nonumber\\
&=16-16\sum_{k_1+k_2+\ldots+k_s=n}{\sigma_1^{2k_1}\sigma_2^{2k_2}\cdot\cdot\cdot\sigma_s^{2k_s}}.
\end{align}

Now we show that for Schmidt rank $s>1$ and in the limit of large $n$ the last sum tends to zero, hence in case of many copies $n$ we get $\FQ[\ket{\psi}^{\otimes n}, \mathcal H_{n-{\rm copy}}]\rightarrow16$. To this end we set $k_1=n-k$ in the last sum above to get the following series of relations:
\begin{align}
&\sum_{k_1+k_2+\ldots+k_s=n}{\sigma_1^{2k_1}\sigma_2^{2k_2}\cdot\cdot\cdot\sigma_s^{2k_s}}\nonumber\\
&\quad\quad\quad\quad=\sum_{k=0}^n\left(\sum_{k_2+\ldots+k_s=k}{\sigma_1^{2(n-k)}\sigma_2^{2k_2}\cdot\cdot\cdot\sigma_s^{2k_s}}\right)\nonumber\\
&\quad\quad\quad\quad=\sigma_1^{2n}\sum_{k=0}^n\left(\sum_{k_2+\ldots+k_s=k}{\sigma_1^{-2k}\sigma_2^{2k_2}\cdot\cdot\cdot\sigma_s^{2k_s}}\right)\nonumber\\
&\quad\quad\quad\quad\le\sigma_1^{2n}\sum_{k=0}^n{\left(\frac{\sigma_2}{\sigma_1}\right)^{2k}\sum_{k_2+\ldots+k_s=k}1},
\end{align}
where the inequality above is due to our assumption $\sigma_2\ge\sigma_k$, in the case of $k>2$. Let us now observe that this last upper bound goes to zero in the case of fixed $s$ and $n$ goes to infinity. This comes from the facts that in that case $\sigma_1^{2n}$ goes to zero, and that $\sum_{k_2+\ldots+k_s=k}1$ is a polynomial function of $s$, hence owing to the Cauchy ratio test the series 
\be
\sum_{k=0}^n{\left(\frac{\sigma_2}{\sigma_1}\right)^{2k}\sum_{k_2+\ldots+k_s=k}1}
\ee
 converges absolutely. $\qed$
 
\section{Maximal metrological gain}

In this section, we consider the multiparticle case. For this case, the metrological gain can be define analogously to the bipartite case.
We determine the quantum states with a maximum metrological gain.

Let us consider the high-dimensional Greenberger-Horne-Zeilinger (GHZ) state 
 \cite{Greenberger1989Going,Krenn2016Automated}
 \be
 \ket{{\rm GHZ}}=\frac 1 {\sqrt{m}} \sum_{n=1}^m \ket{n}^{\otimes N}, \label{eq:hdGHZ}
 \ee
 where $N$ is the number of particles, $d$ is the dimension of their state space, and $m\le d$ is the number of the terms in the superposition.
 We require that $m$ is even.  
 Then, the achievable largest metrological gain 
\be
g( \ket{{\rm GHZ}})=N
\ee
is obtained for the state \eqref{eq:hdGHZ}.
Thus, the maximal gain does not increase with the particle dimension $d$ and depends only on the number of particles. In particular, for two particles, the maximal gain is $2.$

An optimal Hamiltonian with which the maximal gain can be achieved with the GHZ state given in \EQ{eq:hdGHZ} is of the form
 \be
 \mathcal H_{\rm opt}=\sum_{n=1}^N \openone^{\otimes (n-1)} \otimes D' \otimes \openone^{\otimes (N-n-1)} ,
 \ee
 where $\openone^{\otimes 0}=1,$ and the single particle Hamiltonian is defined as 
 \be
D'=\sum_{n=1,3,5,...,m-1} \ketbra{n}-\ketbra{n+1}.
 \ee
Note that for even $d$ and for $m=d,$ the matrix $D'$ equals the matrix $D$ defined in \EQ{eq:Ddef}.
 
In summary, for a given $N$ and $d,$ several of the GHZ states and Hamiltonians  $\mathcal H_{\rm opt}$ give the maximum metrological gain compared to separable states. Note, however, that this does not mean that $\FQ[\ket{\rm GHZ},\mathcal H_{\rm opt}]$ is maximal in all these cases for a given $N$ and $d$. It just means that $\FQ[\ket{\rm GHZ},\mathcal H_{\rm opt}]$ is the largest possible compared to what is achievable by separable states with the same Hamiltonian $\mathcal H_{\rm opt}.$
\vspace{-0.2cm}

\section{Alternative optimization method}

We present a simple alternative of the two-step iterative optimization method of the paper. 
We use the following finding proved in the main text. If we determine the optimal $\mathcal H$ for a given $M$ using Observation 2, the eigenvalues of the optimal $\mathcal H_n$  satisfying \EQ{eq:cnHconst} are $\pm c_n.$ 
We assume that $\mathcal H_n$ is of the form \eqref{eq:Hopt}.
 We set  $\tilde D_n=c_n{\rm diag}(+1,+1,...,+1,-1,-1,...,-1)$ and then vary $U_n$ in order to get the maximal $\FQ(\varrho,\mathcal H_1 \otimes \eins + \eins \otimes \mathcal H_2).$
 
\section{Robustness of metrological usefulness}
 
We define the quantum metrological robustness, $p_{\rm m}(\varrho,\varrho_{\rm noise}),$  where $\varrho$ is some quantum state and $\varrho_{\rm noise}$ a state representing noise. We call $p_{\rm m}$ is the largest noise fraction  $p$ for which the noisy state
\be
\varrho_{p}=p\varrho_{\rm noise}+(1-p)\varrho
\ee
have $g(\varrho_{p})\ge 1$ \cite{Toth2018Quantum}. The bound in \EQ{eq:rphsi+} for noisy maximally entangled states can be formulated for $d=3$ as
\be
p_{\rm m}(\Psi^{({\rm me})},\openone/d^2)=\frac{25-\sqrt{177}}{32}\approx 0.3655.
\ee

In practice, the noise state can be the white noise and $\varrho_{\rm noise}\propto \openone.$ We can also consider an optimization 
\be
\min_{\varrho_{\rm noise}\in S_{\rm noise}}p_{\rm m}(\varrho,\varrho_{\rm noise}),
\ee
which gives the noise tolerance against certain types of noise defined by the set $S_{\rm noise}.$ For instance, the  $S_{\rm noise}$ can contain all states that are metrologically note useful, i.e., for which $g\le 1.$

We can choose another type or parametrization usual in entanglement theory. Given a state $\varrho$ and a metrologically useless state $\varrho_{\rm noise},$ we can call metrological robustness of $\varrho$ relative to $\varrho_{\rm noise},$ the minimal $s\ge 0$ for which
\be
R_{\rm m}(\varrho\vert\vert\varrho_{\rm noise})=\frac{1}{1+s}(\varrho+s\varrho_{\rm noise})
\ee
is useless for metrology.

The robustness can be obtained with a numerical search for the noise fraction for which $g=1.$ We used a search based on interval halving. That is, we start with an interval given by two noise fractions values $p_{\rm L}$ and $p_{\rm H}$ such that $g(\varrho_{p_{\rm L}})\le 1 \le g(\varrho_{p_{\rm H}})$. We test the noise fraction corresponding to the center of the interval. Depending on whether for that noise value $g\ge 1$ or $g<1,$ we reset the lower or the upper boundary of the interval to the center. We repeat this procedure until the size of the interval is sufficiently small. We used a similar procedure to obtain the noise bounds for states with an extra ancilla and two copies of noise states.

We note that there are general relations between the gain-like and robustness-like quantities, that might be used in our case \cite{Uola2019Quantifying,\CITESUPPLABEL}.

\section{Witnessing dimension}

We can use our approach to witness the dimension of the quantum system \cite{Bowles2014Certifying,Brunner2008Testing,Gallego2010Device,Navascues2015Bounding}, or in general, the type of the interaction that is present.
For instance, we can consider the two-qubit singlet state mixed with $p=0.3596$ white noise, see \EQ{eq:plimit}. Such a state is not more useful than separable states, under any Hamiltonian. Thus, 
\be
\max_{{\rm local }\mathcal H}\FQ[\varrho,\mathcal H]\le \FQ^{{(\rm sep)}}.
\ee
If we find that the quantum state is more useful than separable states then it must be connected to an ancilla or a second copy or activated by another quantum state.

Next, we show how to obtain the bound for product states by measurement. We have to create random pure product states $\varrho.$ Then, we can use that
\cite{Helstrom1976Quantum,*Holevo1982Probabilistic,*Braunstein1994Statistical,
*Petz2008Quantum,*Braunstein1996Generalized,Giovannetti2004Quantum-Enhanced,*Demkowicz-Dobrzanski2014Quantum,*Pezze2014Quantum,*Toth2014Quantum,Pezze2018Quantum,Paris2009QUANTUM}
\be
\FQ[\varrho,\mathcal H] =1-F(\varrho,\varrho_{t})t^2/2+O(t^3),
\ee
where $O(t^3)$ respresents terms that are at least third order in $t,$ $F(\varrho,\varrho_{t})$ is the fidelity between the initial state $\varrho$ and the evolved state is
\be
\varrho_{t}=e^{-i\mathcal H t}\varrho e^{+i\mathcal H t}.
\ee
Thus, for a short time evolution, i.e., for small $t$ we have
\be
\FQ[\varrho,\mathcal H] \approx 1-F(\varrho,\varrho_{t})t^2/2.
\ee
Since both of these states are pure product states and we know $\varrho,$ we can measure the fidelity, and use it to measure $\FQ.$ We can even look for the product state that maximizes $\FQ[\varrho,\mathcal H]$ by some search algorithm.

We can also test whether the metrological performance is consistent with some particular interaction. We can compute the maximum for Hamiltonians of the form
\be
\mathcal H_a \mathcal H_A + \mathcal H_B.
\ee
If the metrological performance is better than this maximum, then the form must be different, i.e., there might be two interaction terms between subsystem $A$ and the ancilla "$a$".
 \be
\mathcal H_a \mathcal H_A + \mathcal H_a' \mathcal H_A' + \mathcal H_B.
\ee
Using ideas similar to the ones in our paper, with our method we can even look for the maximum for such Hamiltonians. If the metrological performance is better than this maximum, the interaction between
$A$ and $a$ must contain at least three terms.


\end{document}